\documentclass{aa}  

\usepackage{graphicx}
\usepackage{txfonts}
\usepackage{lipsum}
\usepackage{subcaption}         % necessary for continued figures, example in section 3
\usepackage{lscape}             % to rotate a single page table, example in appendix.
\usepackage{placeins}           % useful with \FloatBarrier, to keep 
\usepackage{defs}
\usepackage{soul}
\usepackage{bbding}
\usepackage[breaklinks, colorlinks, citecolor=blue]{hyperref}
\usepackage{xcolor}

\begin{document}
 
   \title{J1248+4826: A diffuse radio source consistent with re-energized fossil plasma in a galaxy group}

%   \subtitle{Subtitle}

%%%%%%%%%%%%%%%%%%%%%%%%%%%%%%%%%%%%%%%%
% Please separate each author with the \and command
%
% Use the \corrauth to provide the corresponding
% author address. It will be automatically inserted as 
% footnote in the PDF output.
%
% Please DO NOT include ORCIDs next to author names.
% Instead, please provide an active address for each coauthor:
% it will be automatically extracted by EDPS editorial system, 
% and co-authors will be be able to authenticate their ORCID.
%
% Only authenticated ORCIDs will be taken into account.
% ORCIDs included here will be removed.
%%%%%%%%%%%%%%%%%%%%%%%%%%%%%%%%%%%%%%%%

   \author{M. Polletta\inst{1}\corrauth{maria.polletta@inaf.it}
        \and A.~L. Coil\inst{2}\email{acoil@ucsd.edu}
        \and B.~L. Frye\inst{3}\email{bfrye@arizona.edu}
        \and H.~Dole\inst{4}\email{Herve.Dole@ias.u-psud.fr}
%        \and others\inst{3}%\fnmsep\thanks{Shows the usage of elements in the author field}
        }

   \institute{INAF - Istituto di Astrofisica Spaziale e Fisica cosmica (IASF), via A. Corti 12, 20133 Milan, Italy
            \and Department of Astronomy and Astrophysics, University of California, San Diego, La Jolla, CA 92093, USA
            \and Department of Astronomy/Steward Observatory, University of Arizona, 933 N Cherry Ave, Tucson, AZ, 85721-0009, USA 
            \and Universit\'e Paris-Saclay, CNRS, Institut d'astrophysique spatiale, 91405, Orsay, France}

   \date{Received 30 April 2026 / Accepted 2 September 2026}

% \abstract{}{}{}{}{}
% 5 {} token are mandatory
 
  \abstract
% context
{High-resolution, sub-gigahertz radio surveys have revealed a growing number of diverse, diffuse radio sources, presenting new challenges in classifying them and deciphering their nature.}
% aims
{We report the discovery of J1248$+$4826, a complex diffuse radio source in the Low Frequency Array (LOFAR) Two Metre Sky Survey and investigate its physical origin.}
% methods
{We analyzed the radio morphology and spectrum of J1248$+$4826 and studied its environment and optical counterparts using multiwavelength data. After subtracting the radio emission of the bright radio galaxies in the field to isolate the diffuse component, we compared it with known diffuse radio sources including odd radio circles (ORCs), diffuse circular sources, cluster halos, and remnant lobes to establish its nature and formation mechanism.}
% results
{The apparent ring-like morphology, high radio luminosity, and steep spectrum of J1248$+$4826 initially suggested that this source might be a new ORC or a diffuse circular source. However, the residual radio emission reveals a compact ($\sim$30\,kpc radius) horseshoe-shaped inner structure, embedded in an extended ($\sim$200\,kpc) diffuse envelope, with an ultra-steep spectrum ($\alpha=1.7\pm0.2$), which we interpret as a re-energized remnant lobe. The analysis of the multiwavelength data reveals a possible association with a galaxy group at $z=0.2$ and points to its most massive known member, a currently inactive early type galaxy located $\sim$55\,kpc from the structure's center, as the progenitor. Moderate shocks associated with the group dynamics provide a plausible reacceleration mechanism, although the details of such a process remain uncertain with the current data.}
% conclusions
{J1248+4826 highlights the increasing diversity of diffuse radio phenomena and the challenges in their study and classification, while also revealing their potential to study the interplay between galaxies and their environment on large scales.}


   \maketitle
\nolinenumbers
%%%%%%%%%%%%%%%%%%%%%%%%%%%%%%%%%%%%%%%%%%%%%%%%%%%%%%%%%%%%%%
\section{Introduction}

%\LEt{General notes: (A) Acronyms - Please remember to spell out acronyms upon first appearance in the abstract and then again upon first use in the main text (this does not necessarily apply to telescopes, instruments, etc. in the abstract). Thereafter the acronym should be used unless it is at the beginning of a sentence. Please make the necessary changes throughout. (B) TEnse - A\&A uses the past tense to describe specific methods used in a paper, and the present tense to describe general methods and the findings of recent papers. Furthermore, the use of the present tense as opposed to the future is advised for generalities and when authors discuss what can be found within their own work. For more details, see Sect. 6 of the language guide https://www.aanda.org/for-authors/language-editing/6-verb-tenses. Please review my edits to ensure this was carried out appropriately throughout your paper and make any further necessary edits.***}

Modern radio facilities now reveal the Universe with a level of detail and aesthetic complexity that rivals traditional optical imagery, surprising us with spectacular and unexpected structures \citep[see e.g.,][]{brienza21,norris21_orc_discovery}. In recent years sensitive, sub-gigahertz radio surveys at arc-second angular resolutions, which map large areas of the sky, have revealed a rich diversity of diffuse radio sources. The largest diffuse radio sources include radio halos, mini-halos, relics, and bridges associated with the intracluster medium (ICM) of galaxy clusters \citep{vanweeren19,botteon18,govoni19} as well as filaments, bubbles, and shells that form in the intragroup medium (IGrM) of galaxy groups \citep{brienza21,brienza25,arnaudova24}.  Other categories of diffuse radio sources directly linked to individual galaxies, such as active galactic nuclei (AGNs) or star-forming galaxies, are the Fanaroff-Riley radio galaxies \citep{fanaroff74}, gently re-energized tails \citep[GReETs,][]{degasperin17}, and galaxies with large-scale ambient radio emission \citep[i.e., GLAREs;][]{gupta25}. Some recently discovered classes of diffuse radio sources with ring or circular morphologies -- typically associated with massive galaxies -- include the so-called odd radio circles \citep[ORCs;][]{norris21_emu,norris21_orc_discovery,norris21_review,koribalski21} and diffuse circular sources \citep{kumari24,kumari24_horseshoe,kumari25,pal26}. There is also an increasing number of diffuse radio sources associated with fossilized plasma previously deposited by an AGN-driven radio jet. These are remnant radio lobes that are either aging and fading or being re-energized and brightened by shocks or turbulence \citep[e.g., phoenixes;][]{mandal20}.\null

The emission of these diffuse radio sources is due to synchrotron radiation from relativistic electrons in the presence of a magnetic field. They are, in general, the result of AGN feedback processes, episodic jet activity, galaxy interactions, and/or large-scale structure dynamics and, hence, they are powerful probes of fundamental processes that contribute to the shape of our Universe. The discovery and diversity of these diffuse radio sources have also motivated theoretical work and simulations with the aim of reproducing their morphologies and energetics and assessing their formation mechanisms \citep[e.g.,][]{enblin03,hoeft08,paul11,brunetti20,lee24,vazza25}.  However, the nature and formation of some diffuse sources remain unconstrained \citep{delain06,brown09,koribalski24_physalis,bulbul24}. \null

Diffuse radio sources can present a variety of panchromatic information. Although these sources are not typically detected at other wavelengths, observations at optical, infrared, and X-ray frequencies are instrumental for investigating their nature. For example, optical and X-ray observations as well as multifrequency radio observations are crucial for identifying galaxies and AGNs, characterizing their environments, and searching for a diffuse hot gas component that could be associated with the observed radio emission. A multiwavelength approach is also necessary to distinguish projection effects that might produce complex and atypical morphologies from physically disconnected sources. The study and characterization of peculiar radio diffuse sources is fundamental to establish whether they represent a previously known type of phenomenon or a new distinct class of objects and will require increasing effort from the community in the coming years.

In this work we present a new, unusual diffuse radio source, designated as J1248$+$4826, that was serendipitously discovered in the LOFAR \citep{vanhaarlem13} Two Metre Sky Survey \citep[LoTSS;][]{shimwell19,shimwell26}. We analyzed its radio morphology, luminosity, and spectral properties and studied its environment and optical counterparts using multiwavelength data. We compared this source with other diffuse radio sources from the literature to establish its nature and with model predictions to assess its formation scenario. \null

The paper is organized as follows: in Sect.~\ref{sec:radio} we describe the source radio properties. In Sect.~\ref{sec:anc_data} we present the available multiwavelength data in the field, and in Sect.~\ref{sec:src_properties} we identify the counterparts of the radio sources and estimate their redshift and main properties using these data. We investigate the possible origin of the diffuse radio emission in the J1248$+$4826 field by comparing its characteristics with those of other radio diffuse sources in Sect.~\ref{sec:comparison}. We discuss some formation scenarios and detail possible ways to test them in Sect.~\ref{sec:discussion}. Our findings are summarized in Sect.~\ref{sec:conclusions}.  Throughout this work we adopt a \citet{chabrier03} initial mass function (IMF) and a flat $\Lambda$ cold dark matter ($\Lambda$CDM) model, with H$_{\mathrm{0}}$\,=\,67.4\,\kms\,Mpc$^{-1}$, $\Omega_{\Lambda}$\,=\,0.685, and $\Omega_{\mathrm{m}}$\,=\,0.315 \citep{planck_cosmo18}.
   \begin{figure*}[h!]
        \centering
        \includegraphics[width=0.33\hsize]{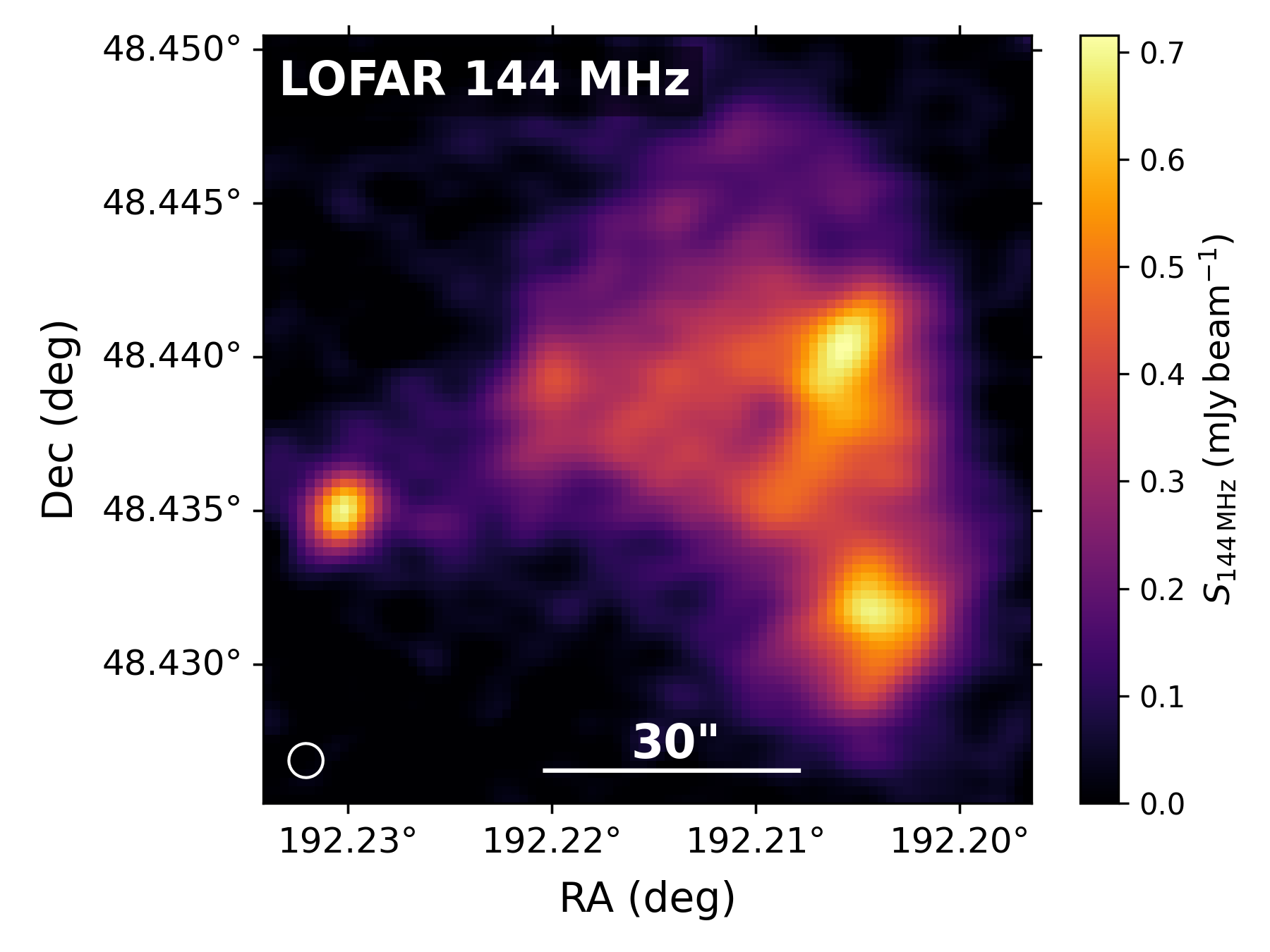}
        \includegraphics[width=0.33\hsize]{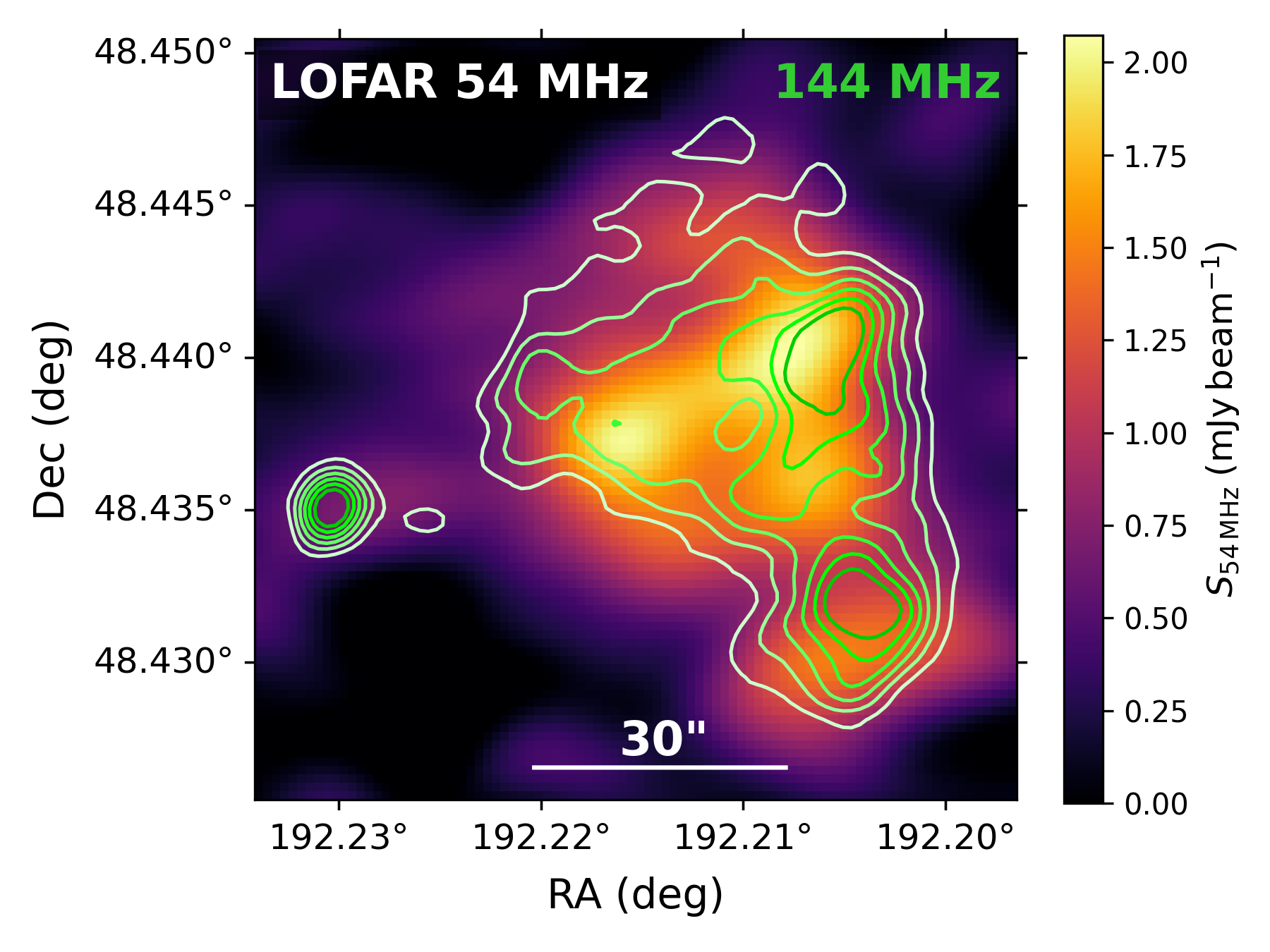}
        \includegraphics[width=0.33\hsize]{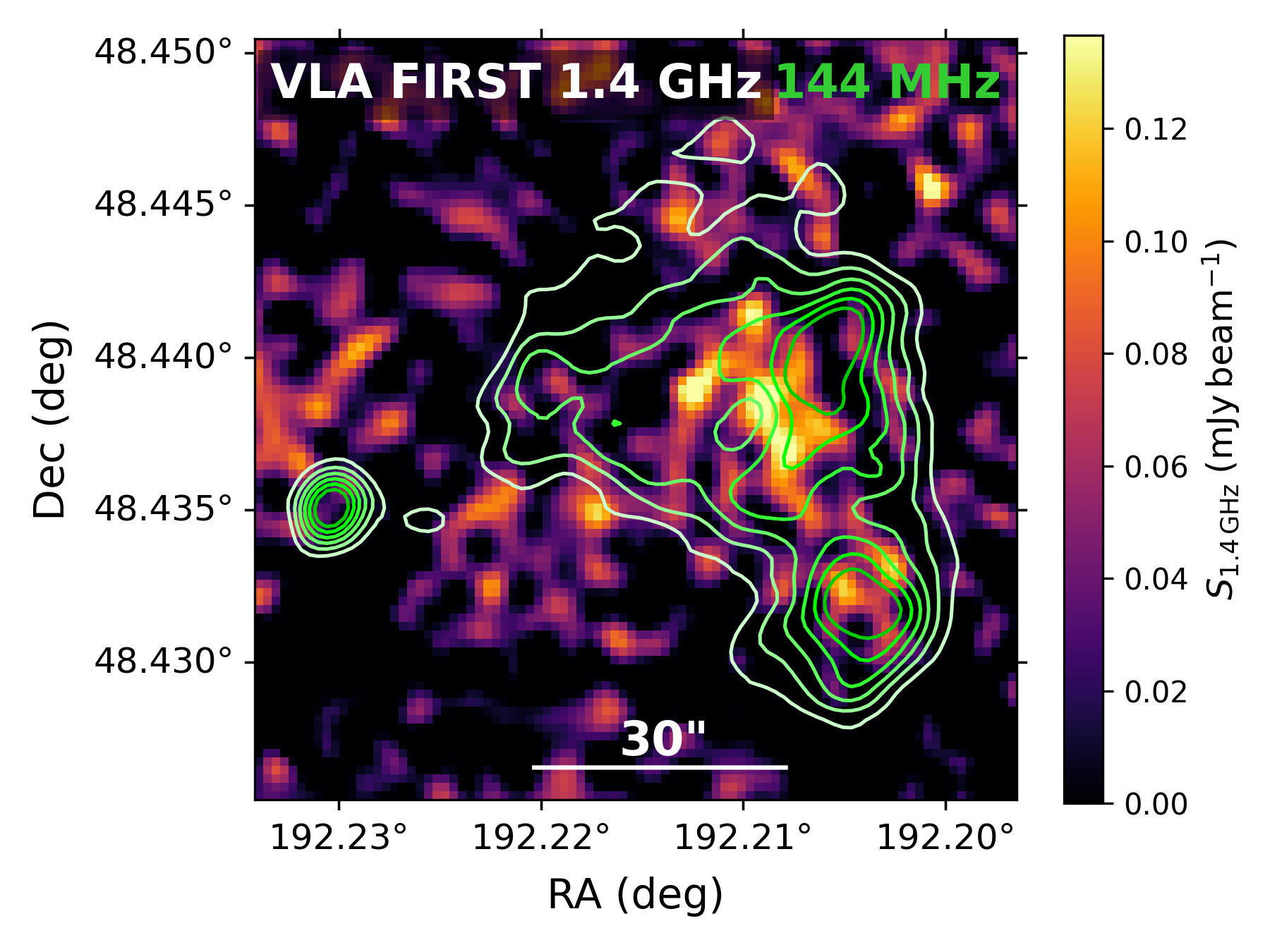}
        \caption{Radio images of the 1.5\arcmin$\times$1.5\arcmin\ field centered
        on J1248$+$4826. \textit{Left:} LoTSS DR3 image at 144\,MHz with 
        a 6\arcsec\ resolution. The white circle in the bottom left
        corner represents the LOFAR beam, a circle with a diameter of
        6\arcsec. \textit{Middle:} LoLSS DR1 image of J1248$+$4826 at 54\,MHz with a 15\arcsec\ resolution. \textit{Right:} VLA FIRST image at
        1.4\,GHz with a 5\arcsec\ resolution. The contours represent the
        intensity of the 144\,MHz LOFAR emission in six steps from 3\,rms to 9\,rms
        from light to dark green with an rms=0.064\,mJy\,beam$^{-1}$. The color scales 
        represent the intensity of the radio emission in millijanskys per beam.
        The horizontal white line represents an angular distance of 30\arcsec.}
        \label{fig:radio_imgs}%
    \end{figure*}
\section{A new diffuse radio source: J1248$+$4826}\label{sec:radio}
We serendipitously discovered an extended and diffuse radio source through a visual inspection of the LoTSS image at 144\,MHz \citep{shimwell19,shimwell26} in proximity of the \planck-selected galaxy protocluster \object{PLCK\,DU\,G124.1+68.8} \citep{planck15,planck16,martinache18,polletta22}. This region is included in mosaic 393 of the LoTSS Data Release 3 (DR3) and in mosaic P190$+$47 of the 54\,MHz LOFAR Low Band Antenna (LBA) Sky Survey (LoLSS) Data Release 1 \citep[DR1;][]{degasperin21}.  The LoTSS observations are characterized by a sensitivity of $\sim$64\,$\mu$Jy\,beam$^{-1}$ (1$\sigma$) and an angular resolution of 6\arcsec, that is the beam full width at half maximum (FWHM). The LoLSS observations reach a sensitivity of 1.76\,mJy\,beam$^{-1}$ (1$\sigma$) at a resolution of 15\arcsec\ (beam FWHM).  This radio source, detected at both frequencies, is characterized by an irregular surface brightness distribution extending across $\gtrsim$1\arcmin, a few moderately bright peaks, and a central ring-like shape with a central depression. We refer to this radio source as J1248$+$4826 (J1248 hereinafter).  The LOFAR images of the J1248 field are shown in Fig.~\ref{fig:radio_imgs}.  The central depression is also visible in the LoLSS image, although it is less obvious because of the coarser angular resolution. 
   \begin{figure}[ht!]
        \centering
        \includegraphics[width=0.95\hsize]{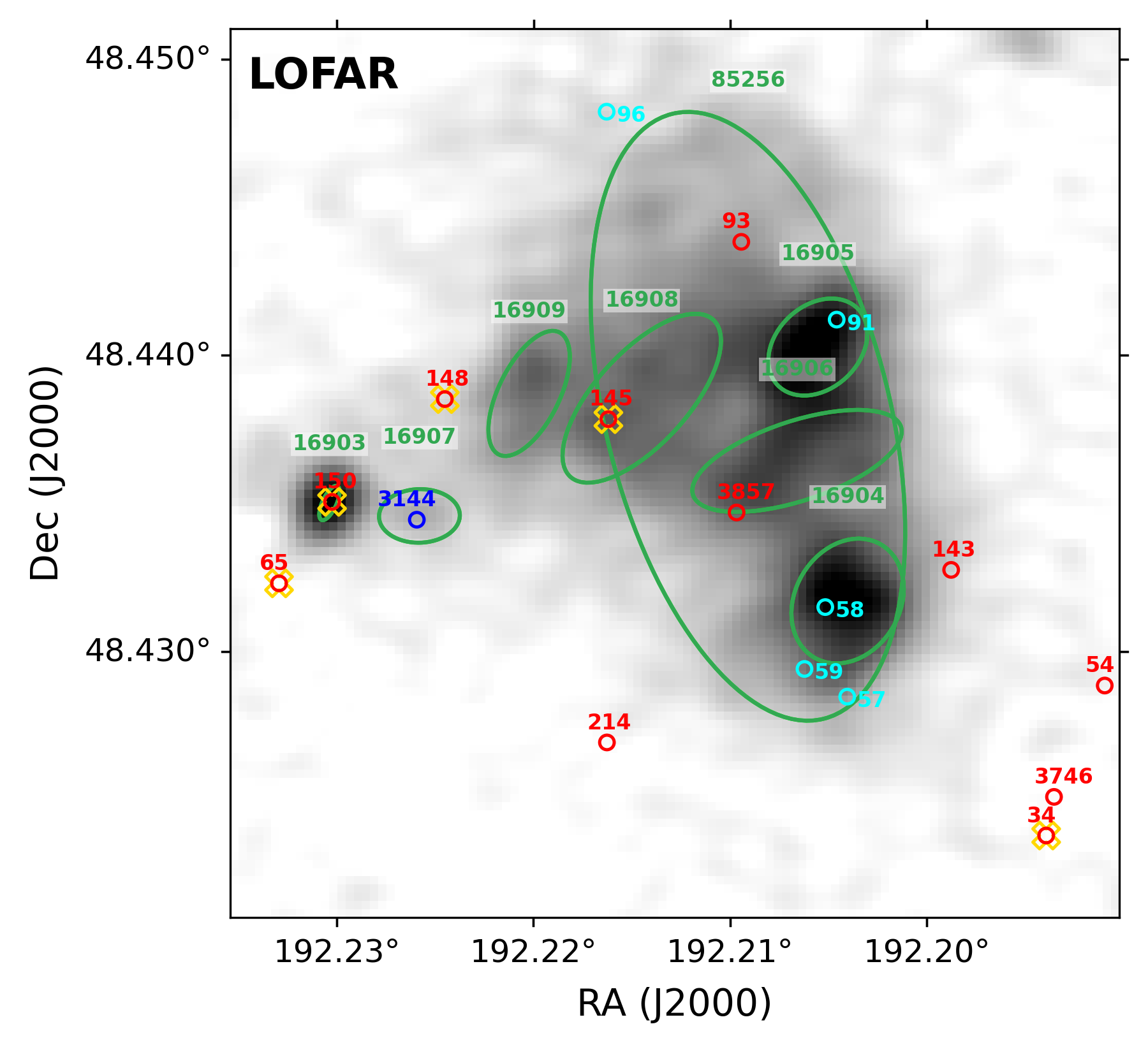}
        \includegraphics[width=0.95\hsize]{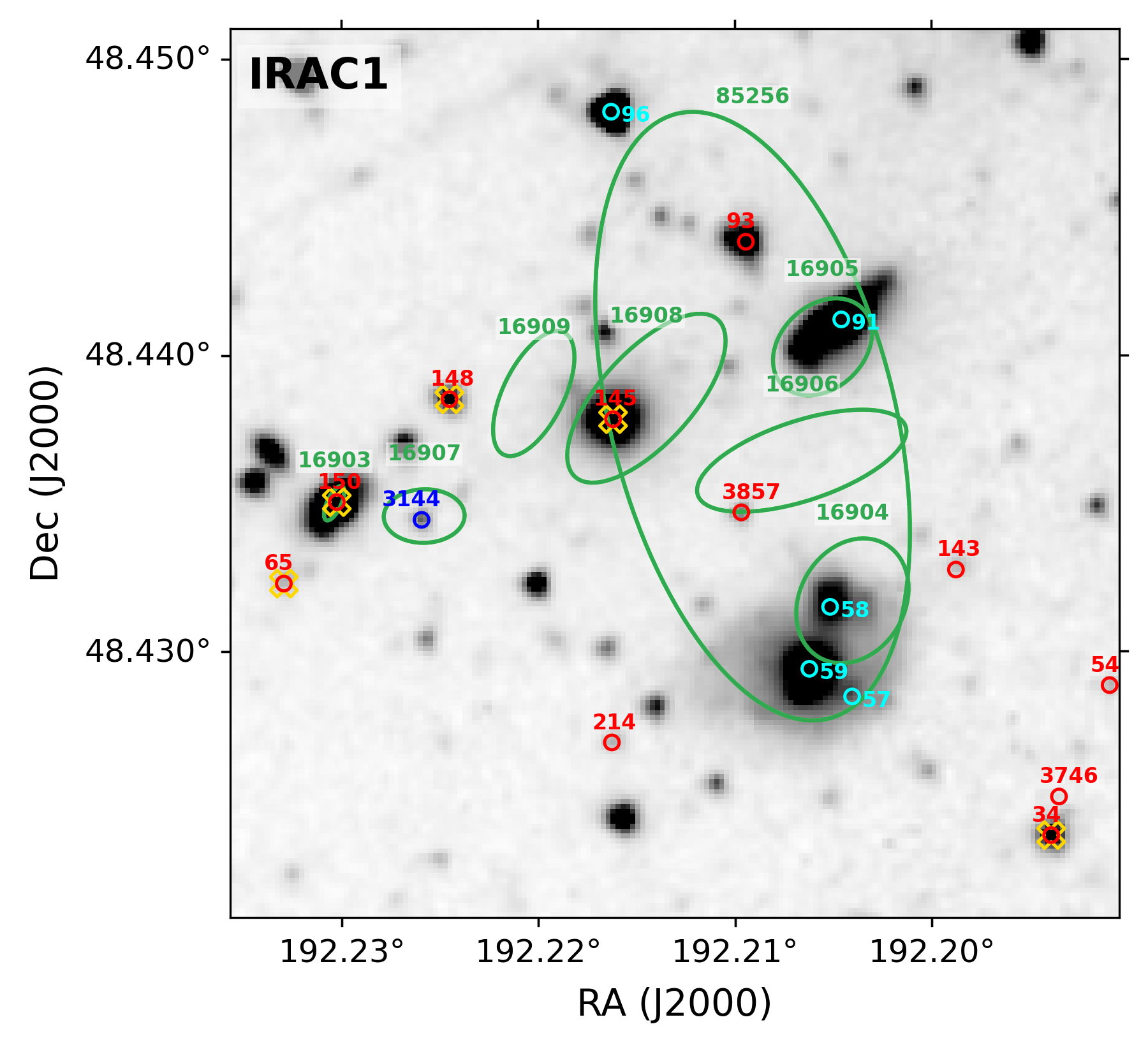}
        \caption{LOFAR 144\,MHz (\textit{top panel}) and IRAC[3.6$\mu$m] (\textit{bottom
        panel}) images of the J1248$+$4826 field. The red
        circles represent the group galaxies at $z\simeq0.2$, the cyan
        circles the galaxies in the foreground ($z<0.1$), and the blue
        circle a mid-IR counterpart to a radio source with no redshift. 
        The green ellipses represent the detected LOFAR sources and their
        extent. The five group members identified by \citet{wen24} and
        \citet{zou22} are marked with yellow crosses. The identification 
        number of each source is marked (see
        Table~\ref{tab:group}).}
        \label{fig:img_wsrcs}
    \end{figure}
\subsection{Radio emission decomposition of J1248$+$4826}\label{sec:radio_props}
There are two sources in the LoTSS DR3 catalog in a 1.5\arcmin$\times$1.5\arcmin\ region centered on J1248$+$4826 and one in the LoLSS catalog.  One of the LoTSS sources is a multiple source that splits into seven Gaussian components after deblending. The list of sources before and after deblending is reported in Table~\ref{tab:lofar_data}, and their positions and extents are shown in Fig.~\ref{fig:img_wsrcs}.  The LOFAR deblended catalog in this region contains four sources (LoTSS IDs 16903, 16904, 16905, 16907) that are likely associated with galaxies and have total flux densities ranging from $\sim$1\,mJy to 8\,mJy, total-to-peak flux ratios of $\sim2-2.9$, and sizes ranging from $\sim$4\arcsec\ to 16\arcsec.  The other four sources do not have a galaxy counterpart (see Sect.~\ref{sec:src_properties}), their emission is extended with total-to-peak flux ratios ranging from 5 to 74, and sizes range from 17\arcsec\ to 76\arcsec.  The radio surface brightness distribution of these eight sources and their misaligned spatial distribution are more consistent with a system made of the superposition of a few radio bright galaxies and diffuse emission rather than with the components of a single radio galaxy.\null
   \begin{figure*}[h!]
        \centering
        \includegraphics[width=0.45\hsize]{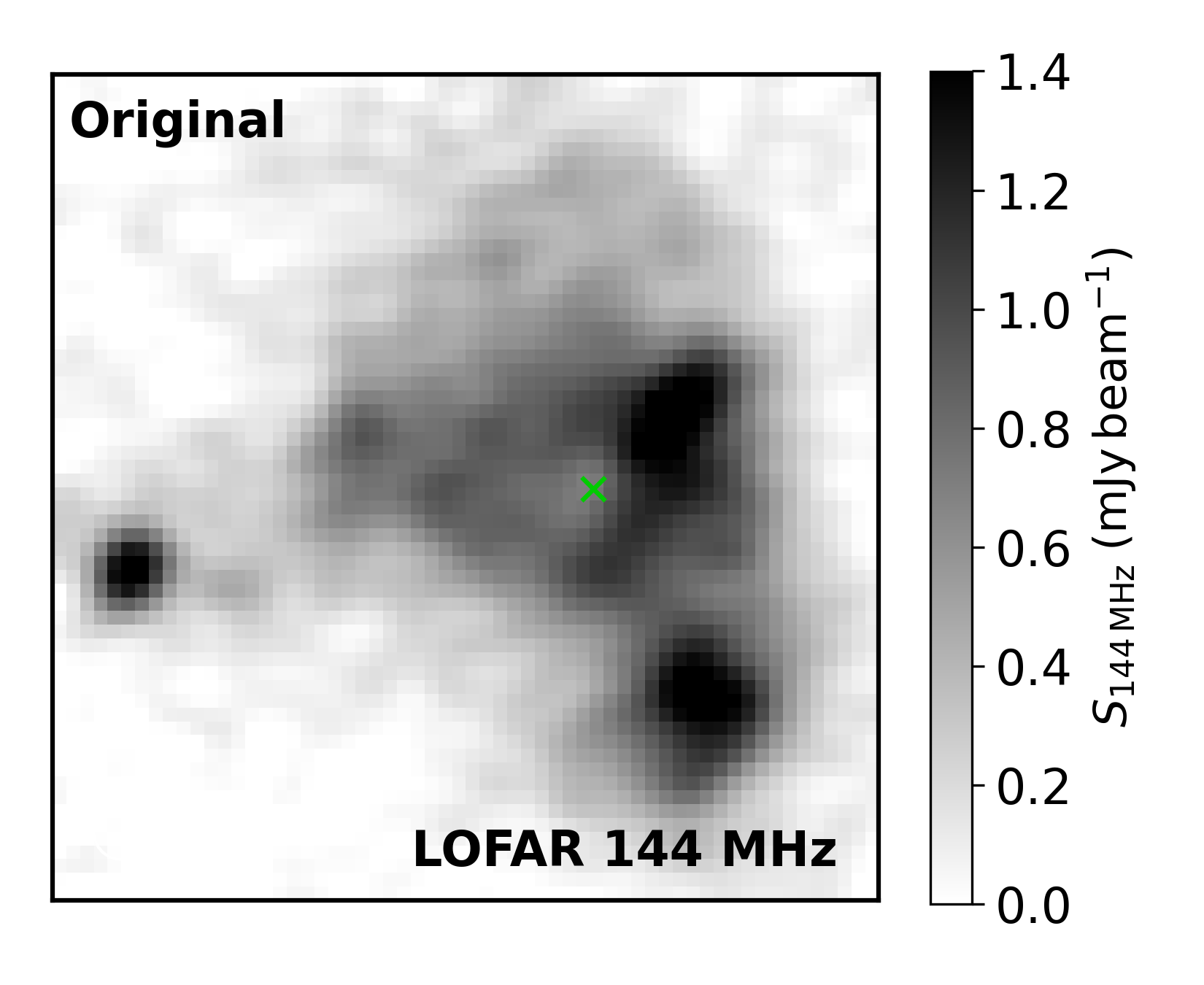}
        \includegraphics[width=0.45\hsize]{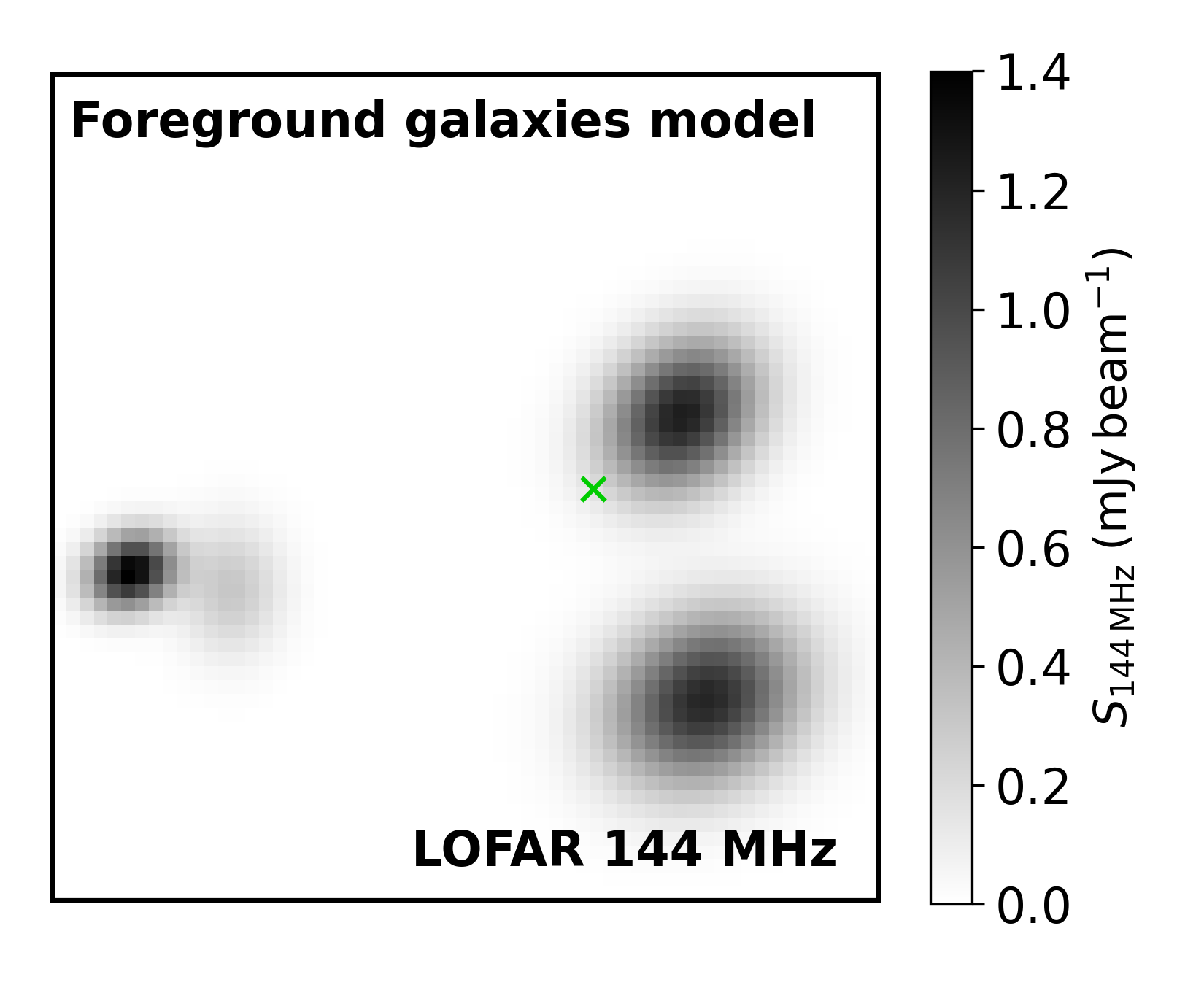}
        \includegraphics[width=0.45\hsize]{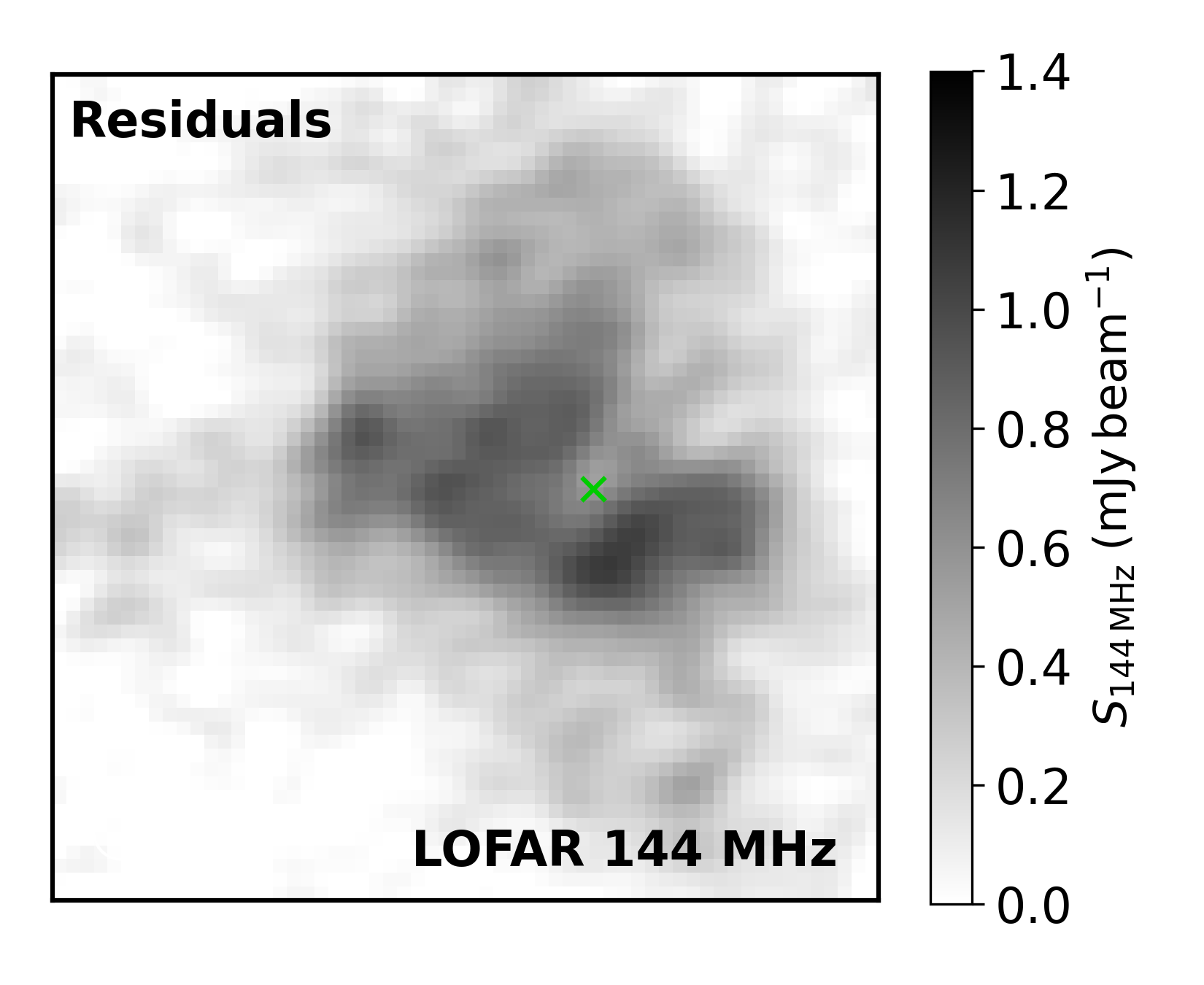}
        \includegraphics[width=0.45\hsize]{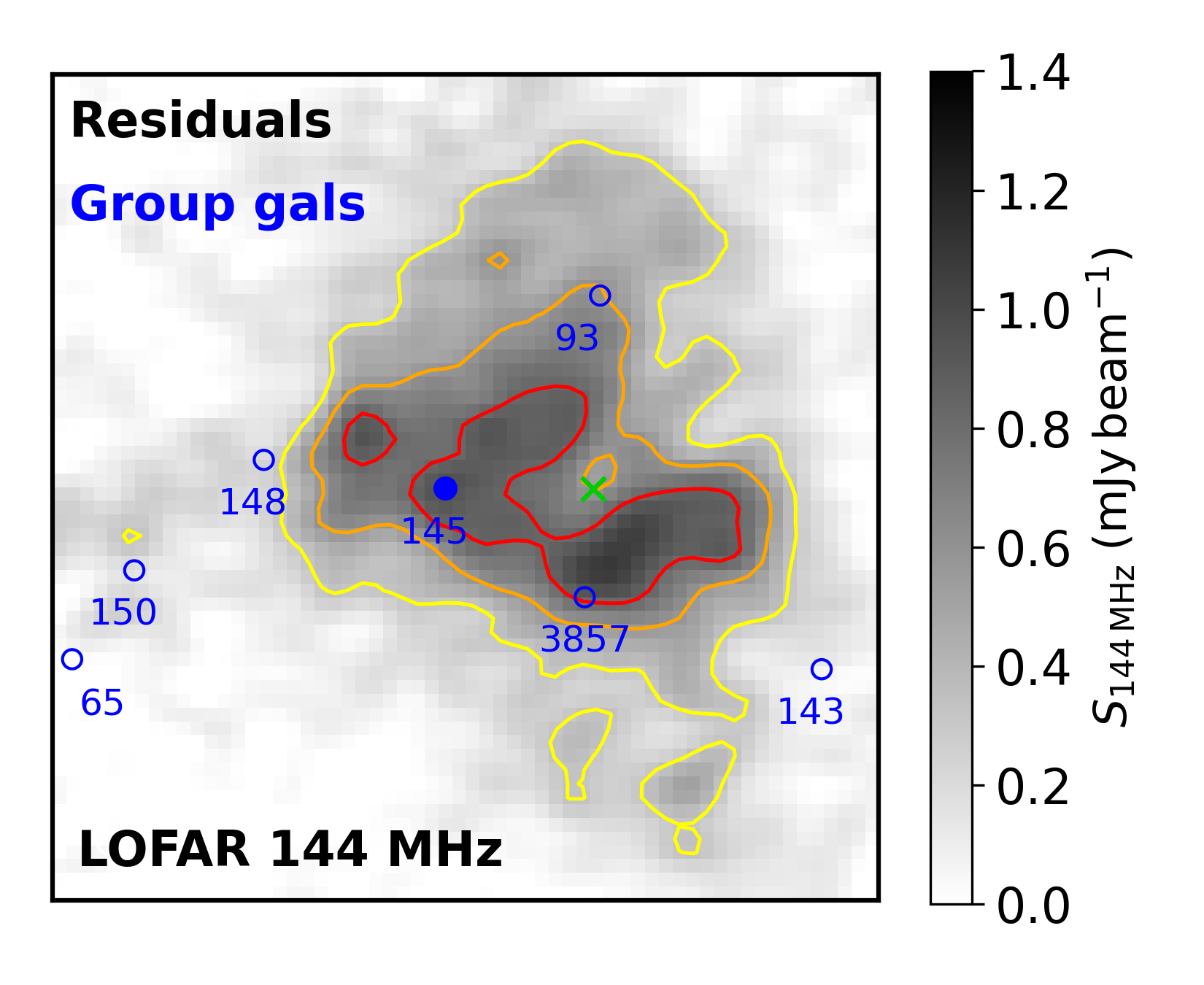}
        \caption{Radio images of the 1.5\arcmin$\times$1.5\arcmin\ field centered
        on J1248$+$4826. \textit{Top left:} LoTSS DR3 image at 144\,MHz. \textit{Top right:} Modeled emission of four radio sources with a galactic counterpart (IDs 16903, 16904, 16905, and 16907; see Table~\ref{tab:lofar_data} and optical counterparts in Table~\ref{tab:radio_lum}). \textit{Bottom left:} Residual radio emission. \textit{Bottom right:} Residual radio emission with the 144\,MHz flux density contours overlaid (yellow, orange, and red correspond to 5, 9, and 13\,rms, respectively) and the $z=0.2$ group member positions (blue circles). The filled blue circle marks the position of ID 145, the putative progenitor of the diffuse emission, and the green cross the center of the radio structure. The gray scale represents the intensity of the radio emission in mJy\,beam$^{-1}$, and it is the same in all panels.}
        \label{fig:radio_residuals}
    \end{figure*}
   \begin{figure*}[ht!]
        \centering
        \includegraphics[height=7cm]{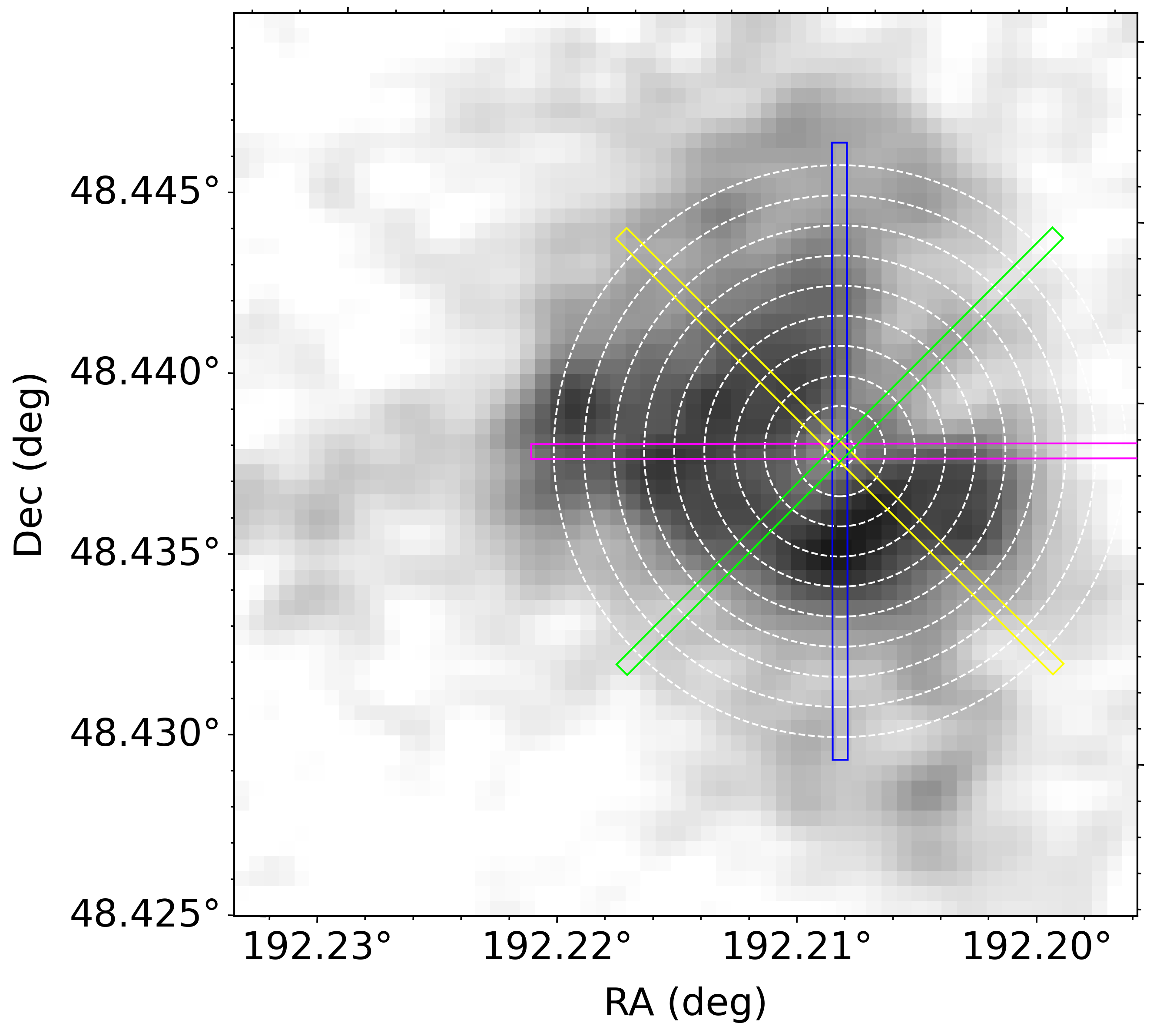}
        \includegraphics[height=7cm]{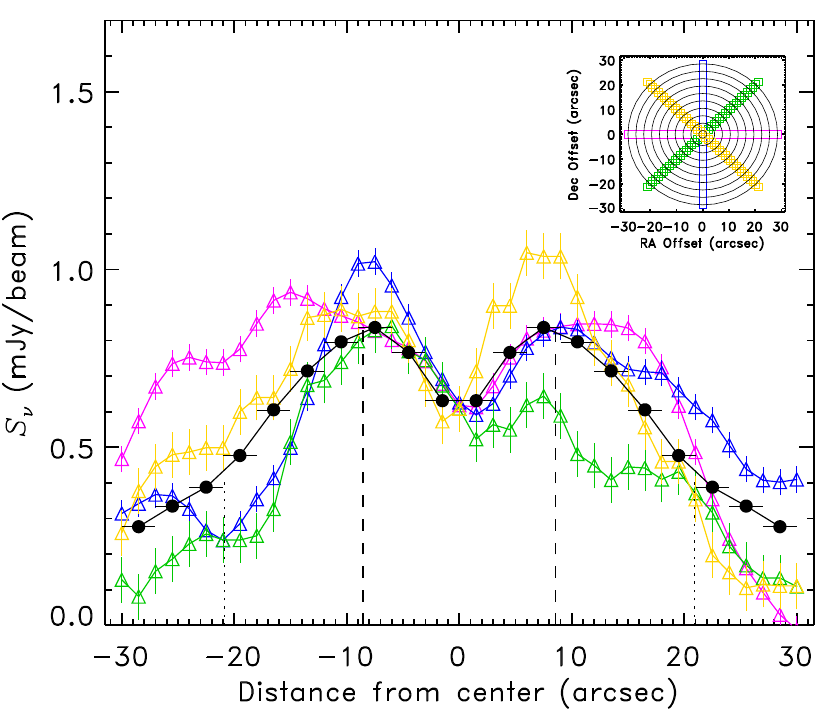}
        \caption{LOFAR residual image and surface brightness profiles in concentric annuli and along four axes centered on the depression and oriented along four different directions offset by 45\deg\ and extending up to 30\arcsec\ from the center on each size.
        The layout of the axes and of the annuli are shown in the left panel and in the inset of the right panel. The flux densities measured after subtracting the contribution from four radio sources with galactic counterparts (LoTSS IDs 16904, 16905, 16907, and 16903) along each axis are shown as triangles in the right panel. The average residual flux density measured in annuli centered on the depression is shown as black circles. The two vertical dashed lines show the radius at which the annular radio profile peaks. The vertical dotted lines represent its FWHM. }
        \label{fig:orc_profile}%
    \end{figure*}
\begin{table*}[ht!]
\caption{\label{tab:lofar_data}LOFAR sources in the J1248$+$4826 field.}
\centering
\setlength{\tabcolsep}{1.8pt}
\renewcommand{\arraystretch}{1.2}
\begin{tabular}{cc cc ccccc}
\hline\hline
 LoTSS  & Source Name               &   $\alpha$ &  $\delta$   & $S^{peak}_{\rm144 MHz}$ & $S^{tot}_{\rm144 MHz}$ & $a_{\rm maj}$ & $a_{\rm min}$ & PA  \\
 ID     &                           &   (deg)    &   (deg)     & (mJy\,beam$^{-1}$)              &  (mJy)                 &   (arcsec)     &  (arcsec)    & (deg) \\ 
\hline
\multicolumn{9}{c}{LoTSS sources (Before deblending)}\\
\hline
        &  ILTJ124850.18+482616.5 &  192.20909  & 48.437939 &  0.43$\pm$0.02 &  31.75$\pm$1.54 &  75.78$\pm$3.69 & 34.5$\pm$1.7 &   14.1$\pm$2.2 \\
        &  \0ILTJ124850.50+482611.2\tablefootmark{a} &  192.21043  & 48.436451 &  1.43$\pm$0.06 &  31.02$\pm$1.00 &  45.85$\pm$2.64 & 30.0$\pm$1.7 &   82.0$\pm$5.3 \\
\hline
\multicolumn{9}{c}{LoTSS Gaussian components (After deblending)}\\
\hline
  85256 &  ILTJ124850.18+482616.5 &  192.20909  & 48.437939 &  0.43$\pm$0.02 &   31.75$\pm$1.54 &  75.78$\pm$3.69 & 34.5$\pm$1.7 & \014.1$\pm$2.2 \\
  16903 &  ILTJ124855.29+482606.0 &  192.23038  & 48.435010 &  1.43$\pm$0.07 &  \01.87$\pm$0.14 & \04.55$\pm$0.39 &\01.7$\pm$0.3 &  155.0$\pm$9.3 \\
  16904 &  ILTJ124848.97+482554.1 &  192.20402  & 48.431703 &  1.20$\pm$0.06 &  \07.97$\pm$0.43 &  15.95$\pm$0.83 & 12.7$\pm$0.7 &  149.1$\pm$9.8 \\
  16905 &  ILTJ124849.33+482625.0 &  192.20553  & 48.440277 &  1.24$\pm$0.06 &  \05.97$\pm$0.36 &  13.47$\pm$0.77 & 10.1$\pm$0.6 &  133.0$\pm$8.6 \\
  16906 &  ILTJ124849.58+482611.1 &  192.20660  & 48.436432 &  0.68$\pm$0.05 &  \05.90$\pm$0.48 &  26.59$\pm$2.20 &\09.7$\pm$0.8 &  108.2$\pm$2.7 \\
  16907 &  ILTJ124854.19+482604.4 &  192.22580  & 48.434575 &  0.31$\pm$0.07 &  \00.89$\pm$0.25 & \09.82$\pm$2.68 &\06.5$\pm$1.8 & \091.5$\pm$31. \\
  16908 &  ILTJ124851.48+482618.7 &  192.21449  & 48.438549 &  0.61$\pm$0.05 &  \05.73$\pm$0.51 &  25.80$\pm$2.30 & 11.2$\pm$1.0 &  137.7$\pm$3.8 \\
  16909 &  ILTJ124852.85+482619.3 &  192.22023  & 48.438711 &  0.58$\pm$0.06 &  \02.70$\pm$0.34 &  16.57$\pm$2.06 &\07.4$\pm$0.9 &  153.7$\pm$5.8 \\
\hline
\multicolumn{9}{c}{LoLSS source} \\
\hline
    185 &  LOL1J124850.3+482614   &  192.20951  & 48.437298 &  15.0$\pm$1.1\tablefootmark{b}  & 286$\pm$21\tablefootmark{b}  &  69$\pm$5 &  59$\pm$4  & 19$\pm$22 \\
\hline
\hline
\end{tabular}\\
\tablefoot{Data from the LoTSS DR3 and LoLSS DR1 source catalogs. The parameters $a_{\rm maj}$, $a_{\rm min}$, and PA are, respectively, the deconvolved FWHM of the source major axis, minor axis, and position angle measured east of north.
\tablefoottext{a}{Source ILTJ124850.50+482611.2 is split in seven Gaussian components in the deblended catalog (from IDs 16903 to 16909).}
\tablefoottext{b}{LoLSS flux density at 54\,MHz.}}
\end{table*}
\subsection{Morphology of the diffuse radio emission}\label{sec:radio_profile}
To examine the morphology of the diffuse radio emission, we subtracted the contribution of four radio sources  with a galactic counterpart (i.e., LoTSS IDs 16903, 16904, 16905, and 16907; see Sect.~\ref{sec:radio_cntp}) uncovered by the deblending procedure and examined the residual image. To do this, we modeled the radio emission of each source as an elliptical Gaussian with the parameters (position, FWHM of the major and minor axis, and position angle) given in the LoTSS DR3 deblended catalog before correcting for deconvolution. The initial, model, and residual images are shown in Fig.~\ref{fig:radio_residuals}. The residual diffuse component exhibits a horseshoe-shaped inner structure (HSS) surrounded by an extended ($\gtrsim$1.0\arcmin\ diameter) envelope with lower surface brightness. The residual diffuse emission amounts to about 73 percent of the total radio flux.\null

We analyzed the residual diffuse emission surface brightness distribution by computing the average flux density profile in concentric annuli and along four radial axes in the residual image.  The annuli and the radial axes are centered at a position estimated to be the center of the HSS ($\alpha=192.20946$\deg\ and $\delta=48.437968$\deg).  The radial axis profiles were obtained by taking the average flux in three pixels across the axis and in steps of one pixel (1.5\arcsec).  Each axis has a length of 41 pixels (61.5\arcsec).  The circular profile is derived by taking the average flux in ten concentric annuli of two pixels width (3\arcsec).  The layout of the four axes and of the annuli is displayed on the left panel of Fig.~\ref{fig:orc_profile}, and the measured flux density profiles are shown in the right panel.  The circular profile is shown twice mirrored in the figure.  The background was not subtracted.  All profiles are characterized by a depression at the center, a peak at 7--15\arcsec\ from the center, and a gradual decrease at larger distances. The circular profile has a flux density peak at $\sim$9\arcsec\ from the center, and a width, computed as the FWHM, of $\sim$21\arcsec.  
Characterizing the exact morphology of the residual emission is challenging due to the uncertainties inherent in bright galaxy subtraction. Notably, using the galaxy models from the LoTSS Data Release 2 (DR2) catalog yields a complete ring rather than a horseshoe, suggesting the "gap" in the structure may be an artifact of the specific subtraction model applied.

\subsection{J1248$+$4826 radio spectrum}\label{sec:radio_spectrum}
\begin{figure}[h!]
        \centering
        \includegraphics[width=\hsize]{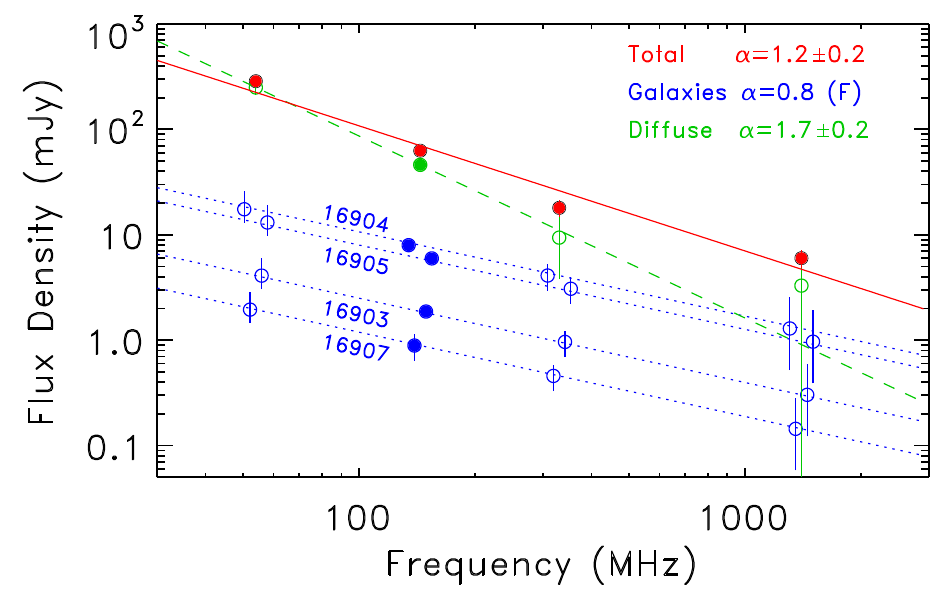}
        \caption{Radio spectra of the whole J1248$+$4826 region (filled red circles), of the four galaxies (blue circles offset along the x-axis for clarity) and of the diffuse emission (green circles) obtained by combining data from the following datasets: LoLSS at 54\,MHz, LoTSS at 144\,MHz, WENSS at 330\,MHz, and NVSS at 1400\,MHz. The filled circles represent measured values, whereas the open circles were derived by assuming that galaxies have a power-law spectrum with spectral index $\alpha=0.8$ (dotted blue lines) and subtracting the galaxy contribution from the total values to estimate the diffuse emission. The best-fit power-law spectra have a slope $\alpha=1.2\pm0.2$ for the integrated emission (solid red line) and $\alpha=1.7\pm0.2$ for the diffuse emission (dashed green line).}
        \label{fig:powerlaw_fit}%
    \end{figure}

A search through public radio catalogs near the J1248$+$4826 field yields two radio sources: \object{NVSS\,124851+482618} from the NRAO VLA Sky Survey \citep[NVSS;][]{condon98} with a flux density $S_{\rm 1.4 GHz}=6.0\pm1.2$\,mJy, and WN\,1246.5$+$4842, from the Westerbork Northern Sky Survey \citep[WENSS;][]{rengelink97} with $S_{\rm 330\,MHz}=18.0\pm3.2$\,mJy. Both sources are detected in surveys at low angular resolution (the beam FWHM is 45\arcsec\ in NVSS, and 54\arcsec\ in WENSS), making them sensitive to diffuse emission and able to collect the flux over an extended region.  No source was found in public radio catalogs obtained from observations at higher angular resolution (beam FWHM = 2.5--5\arcsec) and hence less sensitive to diffuse emission.  Using the flux densities at 1.4\,GHz and at 330\,MHz of these two sources and the total flux densities of the LoTSS 144\,MHz and LoLSS 54\,MHz sources in the field, it is possible to obtain a radio spectrum for the integrated radio emission of J1248$+$4826.  The multifrequency radio data are consistent with a power law spectrum, as shown in Fig.~\ref{fig:powerlaw_fit}.  The spectral index of the integrated radio emission is $\alpha=1.2\pm0.2$ assuming a power-law spectrum (i.e., $S_{\rm \nu}\propto\nu^{-\alpha}$, where $S_{\rm \nu}$ is the flux density and $\alpha$ the spectral index).\null

To derive an estimate of the spectral index associated with only the diffuse component (HSS and envelope) we computed the total flux density at 144\,MHz of the four galaxies and of the remaining four diffuse sources separately (see Table~\ref{tab:lofar_data}).  We then estimated the galaxies emission at the other frequencies assuming a spectral index $\alpha_{\rm gal}=0.8$ \citep{degasperin18} and derived the diffuse flux density by subtracting this estimate from the integrated values.  The different contributions are shown in Fig.~\ref{fig:powerlaw_fit}.  A power-law fit to the diffuse flux densities yields $\alpha_{\rm diff}=1.7\pm0.2$, consistent with an aged electron population.\null

To better constrain the radio spectrum of the diffuse (HSS and surrounding envelope) component and investigate whether there is a spatial gradient across it due to an aging or reacceleration effect, we would need a resolved spectral index map.  Such a map would also allow us to better remove the contribution of the galaxies.  We thus searched for radio images at other frequencies.  We found observations at 1.4\,GHz from the Karl G.  Jansky Very Large Array (VLA) FIRST Survey \citep[rms=0.13\,mJy\,beam$^{-1}$; beam FWHM=5.4\arcsec;][]{becker95}, and the NVSS \citep[rms=0.29\,mJy\,beam$^{-1}$; beam FWHM=45\arcsec;][]{condon98}, and at 3.0\,GHz from the Very Large Array Sky Survey \citep[VLASS;][rms=0.070\,mJy\,beam$^{-1}$; beam FWHM=2.5\arcsec]{lacy20}. The source is unresolved in the NVSS image, undetected in the VLASS image, and barely visible in the VLA FIRST image. The LOFAR and the VLA FIRST radio images of a 1.5\arcmin\ region centered on J1248 are shown in Fig.~\ref{fig:radio_imgs}.  To combine all these maps we would need to reimage the data with the same visibility range and spatially align them with the same restoring beam \citep[see e.g.,][]{biava21}. We will explore the feasibility of such an analysis for a future work, but their shallow depth would probably produce maps with insufficient signal across J1248 to reveal significant spectral variations.

\section{Ancillary data}\label{sec:anc_data}
\begin{figure*}[h!]
        \centering
        \includegraphics[width=0.24\hsize]{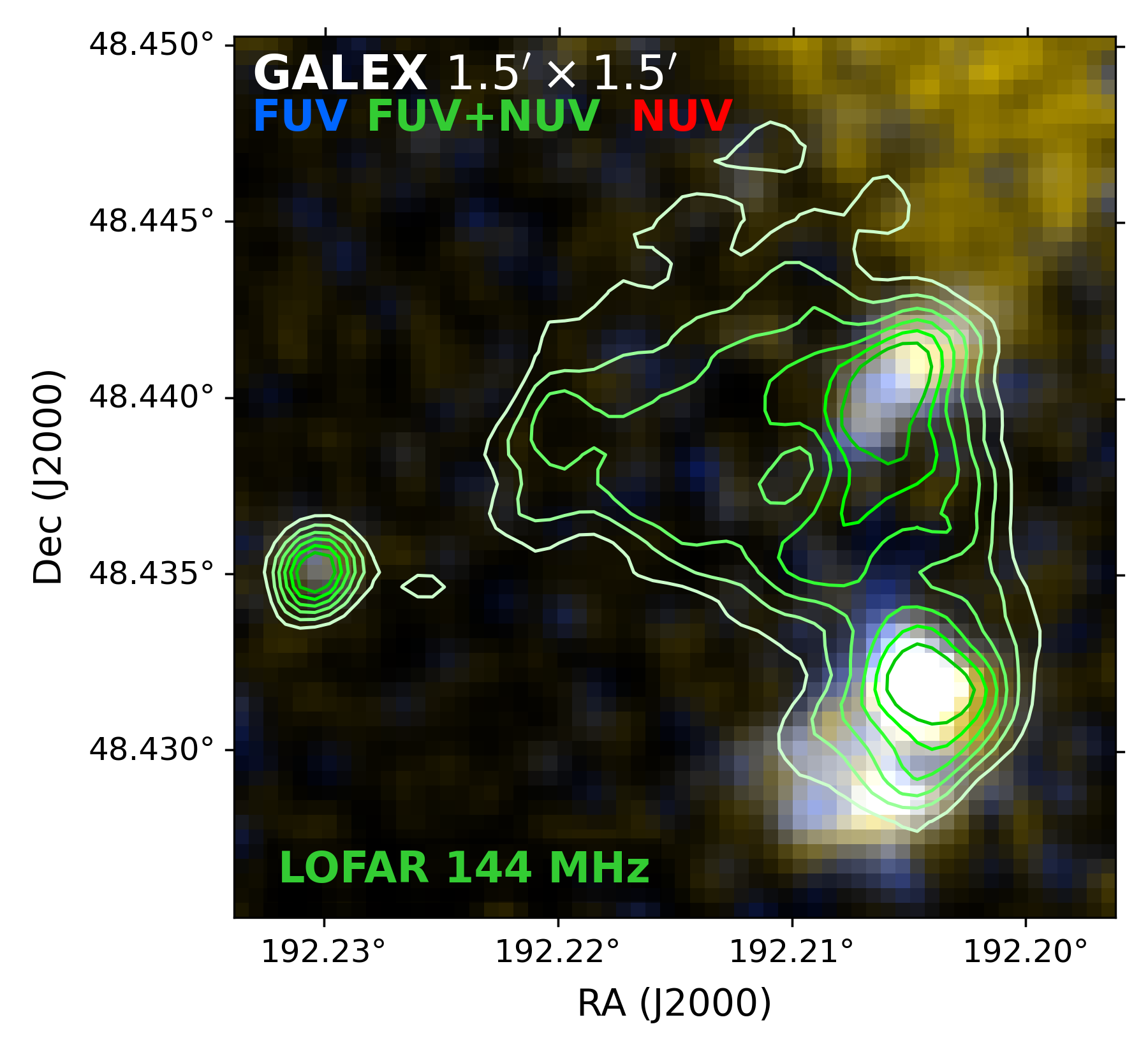}
        \includegraphics[width=0.24\hsize]{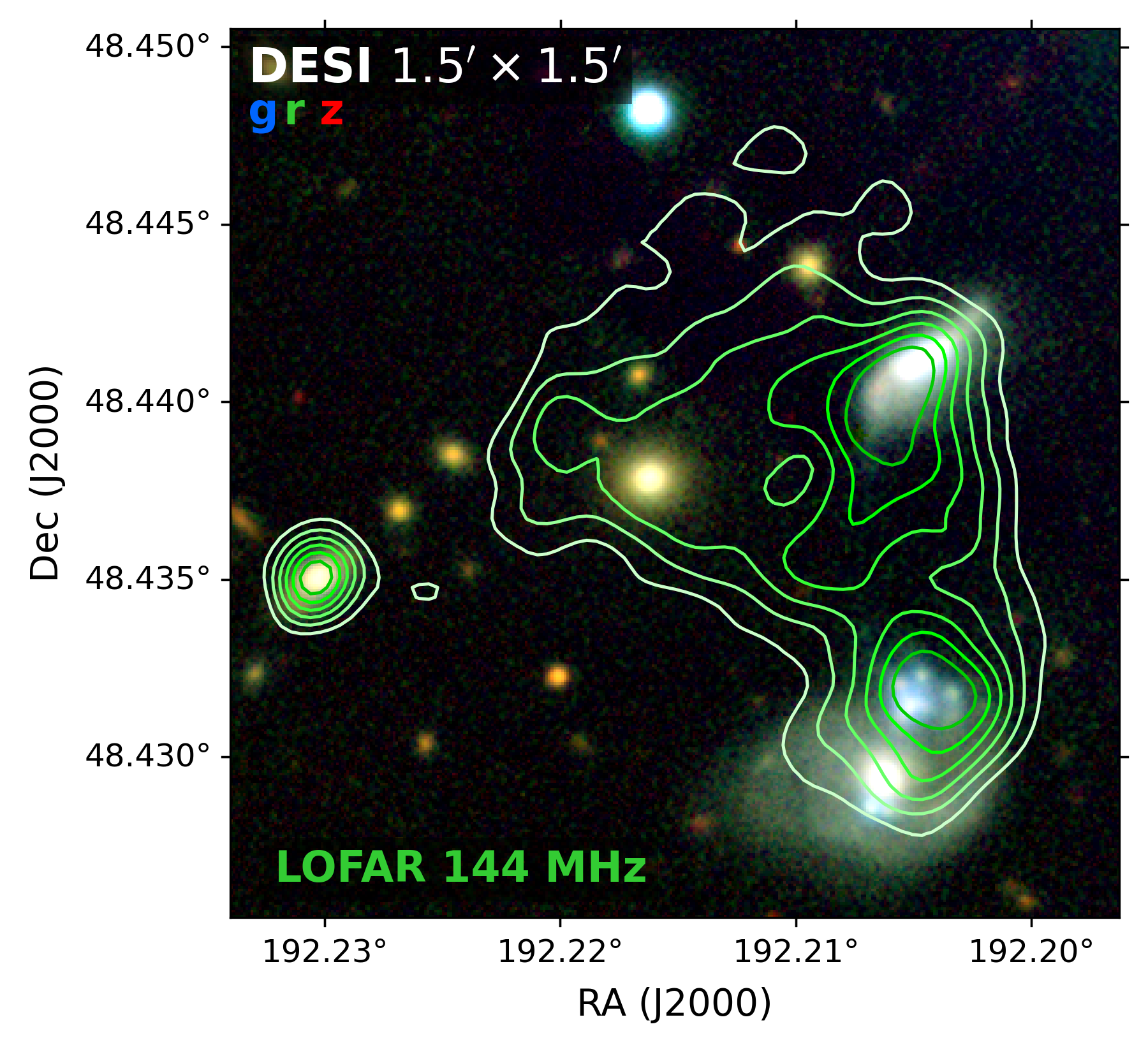}
        \includegraphics[width=0.24\hsize]{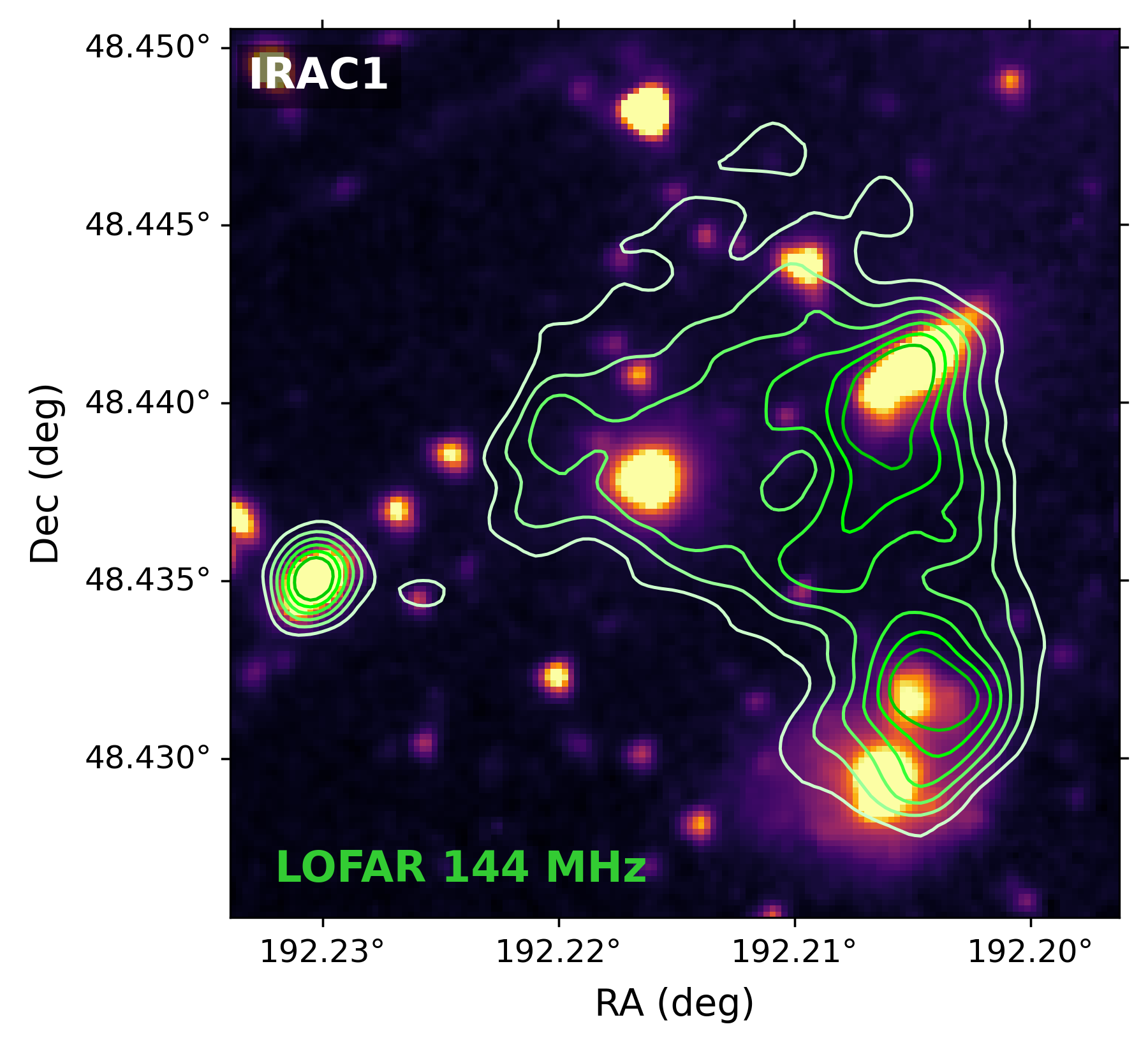}
        \includegraphics[width=0.24\hsize]{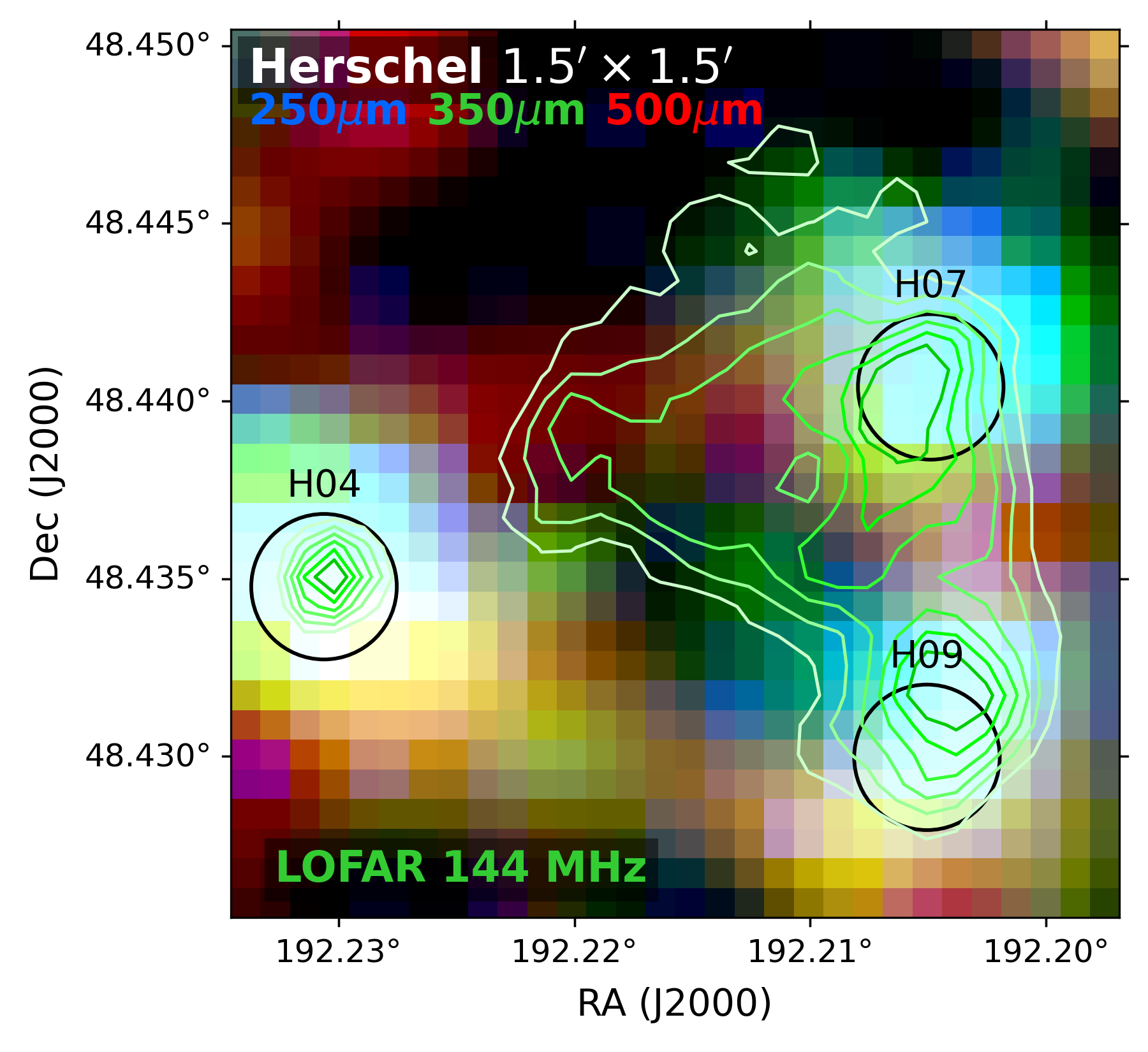}
        \caption{Multiband 1.5\arcmin$\times$1.5\arcmin\ images of the J1248 field. From left to right: UV (GALEX), visible (DESI), mid-IR
        (IRAC1), and submm (\herschel) images.  The bands imaged in the red, green, and blue channels of each multiband image are noted on the top left of each panel.  The green contours represent the LOFAR emission at 144\,MHz as in Fig.~\ref{fig:radio_imgs}.  The black circles and labels in the right panel mark the position and IDs of the \herschel\ sources in the field. }
        \label{fig:multiband_images}%
    \end{figure*}

\begin{figure*}[h!]
        \centering
        \includegraphics[width=0.7\hsize]{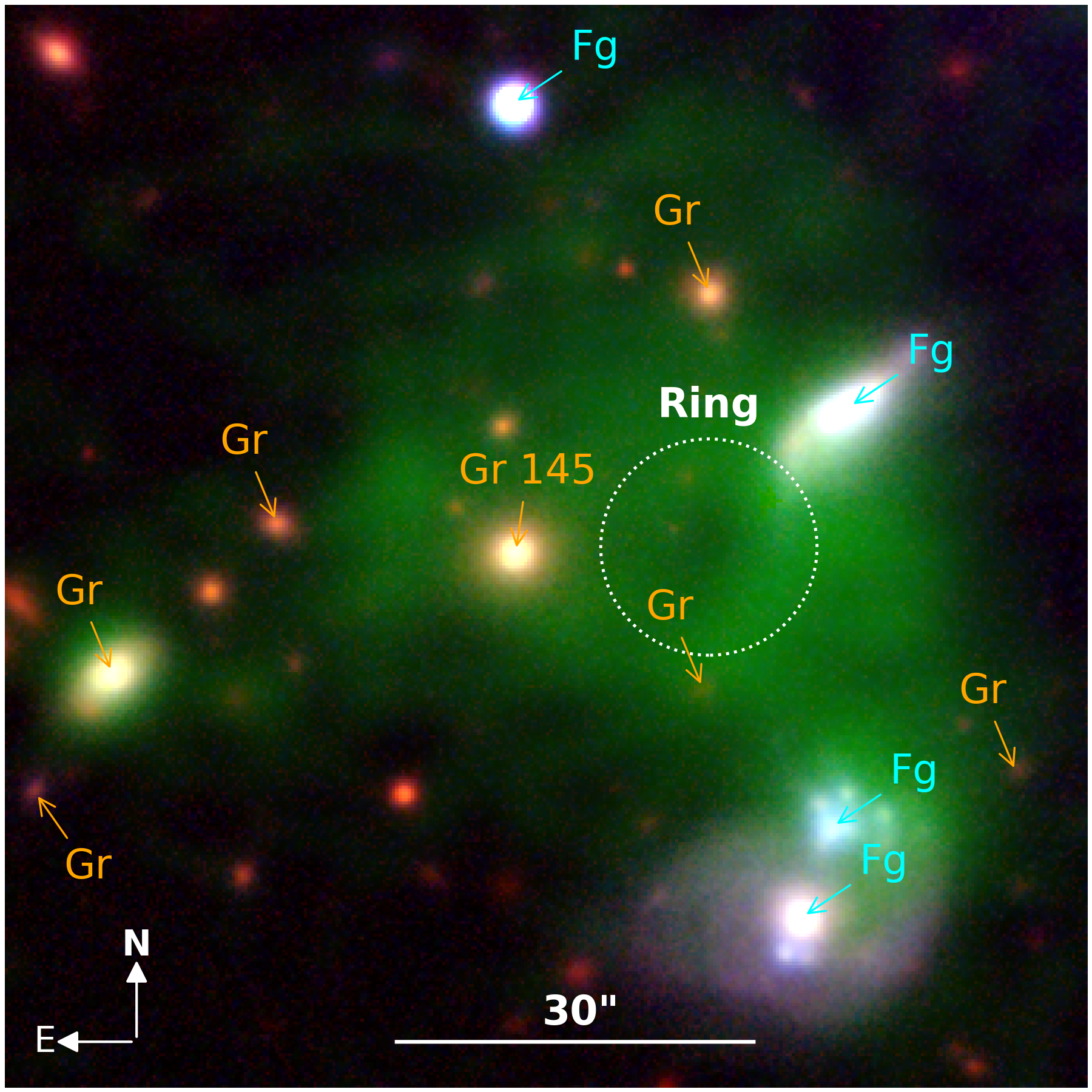}
        \caption{Multiband 1.5\arcmin$\times$1.5\arcmin\ image of the J1248
        field (blue: DESI $g$ band; green: DESI $r$ band; and red: DESI $z$ band
        and IRAC 3.6\,$\mu$m) with the LOFAR 144\,MHz emission overlaid in
        green. The surface brightness peak of the diffuse emission is marked with a dotted white circle with a 9\arcsec radius and labeled "Ring." The member candidates of a $z=0.2$ galaxy group are indicated with orange arrows and labeled "Gr," and the foreground sources overlapping with the radio emission are indicated with cyan arrows and labeled "Fg." The putative progenitor of the radio diffuse emission is the group member ID 145. The horizontal white segment represents a 30\arcsec\ angular distance that is equivalent to a physical distance of about 103\,kpc at $z=0.2$. }
        \label{fig:radio_on_multilambda}%
    \end{figure*}

To help with the interpretation of J1248, we searched for public multiwavelength data in the field. In the visible, we retrieved images and magnitudes in the $grz$ bands from the Dark Energy Spectroscopic Instrument \citep[DESI\footnote{https://www.legacysurvey.org}; ][]{desi} legacy imaging surveys, the Mayall z-band Legacy Survey \citep[MzLS;][]{mzls}, and the Beijing Arizona Sky Survey \citep[BASS;][]{bass} to a nominal 5$\sigma$ depth of 23.48, 22.87, 22.29 AB magnitudes in the $g$, $r$, and $z$ bands, respectively \citep{dey19}.  Mid-infrared (MIR; 3.6\,$\mu$m) data from the Infrared Array Camera (IRAC) on the {\it Spitzer} Space Telescope are available as part of the observations covering the PLCK\,DU\,G124.1+68.8 field~\citep{martinache18} to a 50\% depth of 2.54$\mu$Jy.  Additional MIR data (at 3.4, 4.6, 12, and 22$\mu$m) were obtained from AllWISE \citep{wright10,mainzer11}.  The region was also observed by the \herschel\ Space Observatory at submillimeter (submm) wavelengths through the program targeting the PLCK\,DU\,G124.1+68.8 field~\citep{planck15} with 1$\sigma$ (instrumental and confusion) noise levels of 9.9\,mJy at 250$\mu$m, 9.3\,mJy at 350$\mu$m, and 10.7\,mJy at 500$\mu$m.  There are three submm sources in the field, and they all have a radio counterpart in LoTSS.  The list of \herschel\ sources, their radio counterpart, and submm flux densities are reported in Table~\ref{tab:herschel}.  Ultraviolet (UV) observations in FUV (1516\AA) and NUV (2267\AA) are also available from GALEX~\citep{martin05}.  Images of the J1248 field from these datasets are shown in Fig.~\ref{fig:multiband_images}. A multiband image with the radio emission superimposed is displayed in Fig.~\ref{fig:radio_on_multilambda}. 
\begin{table}
\caption{\label{tab:herschel}\herschel\ sources in the J1248$+$4826 field.}
\centering
\setlength{\tabcolsep}{1.4pt}
\renewcommand{\arraystretch}{1.2}
\begin{tabular}{cc cc ccccc}
\hline\hline
 \herschel & LoTSS  &   $\alpha_{Herschel}$ &  $\delta_{Herschel}$   & $S_{\rm250 \mu m}$ & $S_{\rm350 \mu m}$ & $S_{\rm500 \mu m}$ \\
 ID        & ID     &   (deg)    &   (deg)     & (mJy)              &  (mJy)             &   (mJy)            \\
\hline
  H04      & 16903  &  192.23062  &    48.434776 &  130$\pm$11 &   79$\pm$10 &   $<$36 \\
  H07      & 16905  &  192.20489  &    48.440402 &  117$\pm$11 &   57$\pm$10 &   $<$36 \\
  H09      & 16904  &  192.20505  &    48.429975 &   74$\pm$15 &   55$\pm$11 &   $<$36 \\
\hline
\hline
\end{tabular}\\
\tablefoot{Upper limits correspond to 3$\sigma$.}
\end{table}
No X-ray data are available in the field \citep[based on a search in the Analysis Center for Extended Data, CADE; ][]{paradis12}, with the exception of ROSAT and e-ROSITA observations. However, no detection is available from the former, and no public data exist for the latter.  We find only one source with a public spectroscopic redshift in the field \citep[\object{WISEA\,J124851.91+482616.0} at $z=0.20807\pm0.00004$, ID 145 hereinafter;][]{sdss_dr6}.  Photometric redshifts are available for several sources in the field through the DESI collaboration \citep{zou22,wen24}.
\section{Source properties in the J1248$+$4826 field}\label{sec:src_properties}
\subsection{Spectral energy distributions}\label{sed:seds}
We created a multiband catalog by combining the data from DESI ($grz$), IRAC[3.6$\mu$m], WISE, \herschel, and LOFAR in a 1.8\arcmin$\times$1.8\arcmin\ region centered on J1248. The selection yielded 90 sources, of which 69 have at least two $>3\sigma$ detections. The different catalogs are matched using the source positions taking the respective position uncertainty into account. The multiband catalog was then used to estimate photometric redshifts, classify the different source types and derive the main physical properties. This step was carried out by modeling the spectral energy distributions (SEDs) with the Code Investigating Galaxy Emission \citep[\texttt{CIGALE};][]{boquien19}.  When a spectroscopic or a DESI photometric redshift \citep{zou22,wen24} were available (for 17 sources), we fixed the redshifts at such values. We adopted the BC03 \citep{bruzual03} stellar population models and chose a star formation history (SFH) modeled with a double exponential with a recent burst with a constant folding time ($\tau_{\rm burst}=20$\,Gyr).  These choices should produce the most accurate stellar masses and SFRs among all those that are possible within \texttt{CIGALE} \citep{michalowski14,osborne24}.  Dust emission was modeled using the models by \citet{draine14} and dust attenuation by the two-component model of \citet[][CF00 hereinafter]{charlot00}.  The CF00 model assumes a different attenuation law and strength for the birth clouds and the interstellar medium (ISM) as it can be the case when the visible-NIR and FIR emissions are not co-spatial \citep[e.g.,][]{chen15,hodge16,smail23}.  An AGN component was included assuming the SKIRTOR models \citep{stalevski16}.  Finally, we included a radio component to model the synchrotron emission produced by the interaction of relativistic electrons from supernovae with the local magnetic field.  This radio component is related to the galaxy star formation activity, and modeled as a power law with spectral index $\alpha_{\rm radio}=0.8$ \citep[$F_{\nu}\propto\nu^{-\alpha_{\rm radio}}$;][]{brienza17,degasperin18}.  Its strength is bound to the estimated SFR through the radio-FIR correlation \citep{helou85}.  The best-fit parameters and associated uncertainties are the likelihood-weighted means and standard deviations, respectively.
\subsection{Counterparts to the radio sources}\label{sec:radio_cntp}
To determine the origin of the radio emission in the J1248 field, we used the multiband catalog to identify the counterparts to the radio sources listed in Table~\ref{tab:lofar_data}.  The position of the sources in the multiband catalog that overlap with the radio emission in the J1248 field are shown in the LoTSS and IRAC1 images in Fig.~\ref{fig:img_wsrcs}.  Through the matching procedure, we identified four galaxies in the multiband catalog that are the counterparts to four radio sources, two at $z<0.1$ (i.e., ID 58 to LoTSS ID 16904, and ID 91 to LoTSS ID 16905), one at $z=0.2$ (ID 150 to LoTSS ID 16903), and a fourth one with no redshift estimate because it is only detected by IRAC at 3.6$\mu$m (ID 3144 to LoTSS ID 16907).  Three galaxies (IDs 58, 91, and 150) are also detected in the submm by \herschel\ (see Table~\ref{tab:herschel}) and in the UV by GALEX (see Fig.~\ref{fig:multiband_images}). These detections indicate that they are star-forming galaxies because massive, young stars are usually bright submm and UV emitters.  Their young stellar populations are also radio emitters and their radio luminosity is correlated with the galaxy star formation rate (SFR).  Since in all three cases the radio flux is consistent with the estimated SFR based on the radio-SFR relation \citep{kennicutt12}, there is no need to invoke an additional radio source such as an AGN to explain their radio brightness.  Regarding the fourth source with a radio counterpart (ID 3144 to LoTSS ID 16907), it is not possible to determine the origin of its radio emission because the galaxy is only detected in one band and no SED model is available.  However, its contribution to the total radio emission is negligible, accounting for only 1.4\% of the total radio flux density in the field.\null

There are four additional radio sources in the field that are extended (major axis $\sim$20\arcsec--80\arcsec) and do not have an obvious counterpart in the multiband catalog.  Interestingly, there are several galaxies in the field that, in part, overlap with this diffuse radio emission and that are members of a galaxy group at $z=0.2$ (red circles in Fig.~\ref{fig:img_wsrcs}).  This galaxy group was previously identified using DESI photometric and spectroscopic redshifts by \citet{wen24},  WHJ124851.9$+$482616 at $z=0.2057$. The group contains six members (four in the J1248 field and two just outside), including ID 150 that is a radio source.  Thanks to additional photometric redshifts from our analysis and from \citet{zou22}, we found seven additional members, all with $0.183<z<0.221$ or |$\Delta v$|$\lesssim5000$\,\kms.  The apparent velocity dispersion inferred from the available redshifts of the group members, $\sigma_{v}=2240$\,\kms, is much higher than expected in galaxy groups, but this value is dominated by the large uncertainty in the photometric redshift estimates.  The estimated redshifts of the 11 group members and their main properties are listed in Table~\ref{tab:group}.  Their SEDs and best-fit models are shown in Fig.~\ref{fig:seds}.\null

\begin{figure}[ht!]
        \centering
        \includegraphics[width=\hsize]{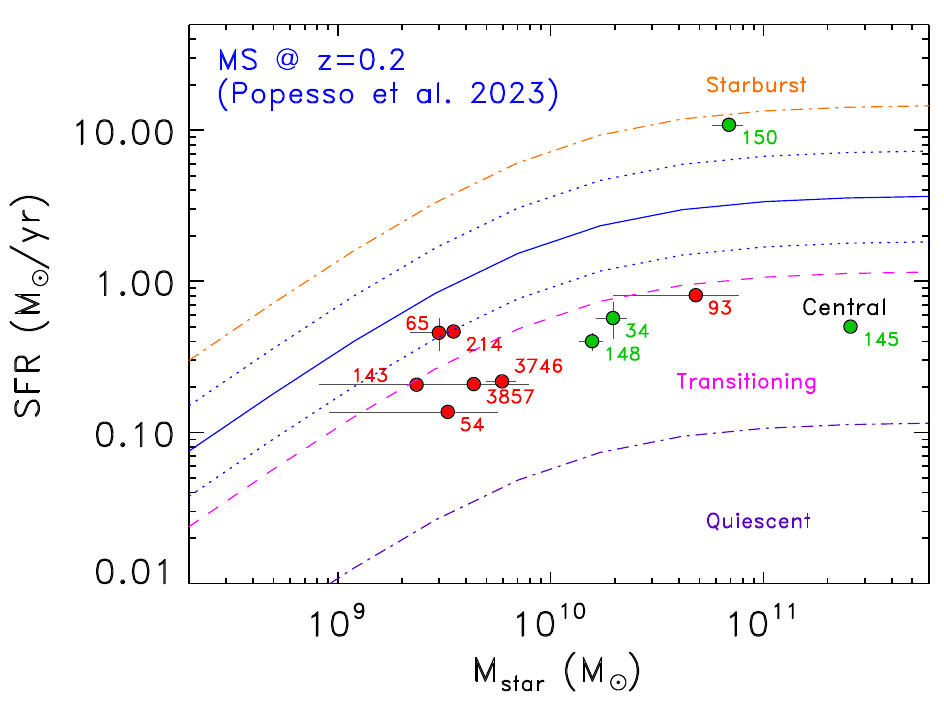}
        \caption{Star-forming MS at $z=0.2$ and its scatter
        \citep[solid and dotted blue lines;][]{popesso23}, along with SFRs and
        stellar masses derived from \texttt{CIGALE} for the 11 group members in
        the J1248 field (filled circles). The green symbols represent the
        group members from \citet{wen24}, and the red symbols represent those identified
        in this work. Sources classified as starburst, transitioning, or
        quiescent are expected, respectively, above the dot-dashed orange curve
        \citep{rodighiero11}, below the dashed magenta curve, and below the
        dot-dashed purple curve \citep{renzini15}. The sources are labeled
        with their identifiers (see Table~\ref{tab:group}).}
        \label{fig:MS}
    \end{figure}

To further validate the group hypothesis we compared the member candidates SFRs with those of the main sequence (MS) of star formation at $z=0.2$.  The population of groups is expected to be dominated by early type galaxies (ETGs) \citep{weinmann06}.  The MS represents the relation between SFR and stellar mass of coeval normal star-forming galaxies \citep[see Fig.~\ref{fig:MS};][]{popesso23}.  All but one of the identified group members are characterized by a low activity level (SFR$\lesssim$1\,\msun\,yr$^{-1}$), and $\sim$70\% of them are situated below the MS, in the zone where "transitioning" galaxies lie \citep[between 0.5 and 1.5\,dex below the MS;][]{renzini15}, consistent with a group population.  \null

Given that diffuse radio sources frequently reside within galaxy groups, we must assess whether the WHJ124851.9+482616 group is associated with J1248 by chance alignment or is a physical association. Assuming a uniform sky distribution of the 1.58 million groups and clusters identified across 24,000\,\sqdeg\ by \citet{wen24}, the probability of finding at least one such system in a random $1.5^{\prime} \times 1.5^{\prime}$ field is about 4\%.  While a chance coincidence cannot be strictly ruled out, this low probability strongly favors a physical connection.  In summary, the total radio emission in the J1248 field arises from four discrete galaxies at $z \sim 0.05 - 0.2$ and, most likely, diffuse plasma within the IGrM of a $z = 0.2$ group.  Notably, this diffuse intragroup component dominates the field's radio flux budget, accounting for $\sim 73\%$ of the total emission.
\subsection{Radio luminosities}\label{sec:radio_lums}
We estimated the radio luminosities at the rest-frame frequencies 144\,MHz and 1.4\,GHz of J1248 by considering the flux density of each radio source, the estimated redshift, and assuming a power-law spectrum defined as $S_{\nu}\propto \nu^{-\alpha}$.  Four radio sources (LoTSS IDs 16903, 16904, 16905, and 16907) are associated with galaxies, while the other four are considered diffuse sources. Of these, two (LoTSS IDs 16906 and 16908) enclose the depression and form the HSS, one (LoTSS ID 85256) traces the low surface brightness extended envelope, and a fourth one (LoTSS ID 16909) is a blob to the east with no galactic counterpart. For the four sources with a galactic counterpart, we adopted the galaxy photometric or spectroscopic redshift, when available, or that of the galaxy group and assumed a spectral index $\alpha=0.8$ \citep{brienza17,degasperin18}.  Several studies report a spectral flattening toward sub-gigahertz frequencies, with the median index dropping to $\alpha\simeq0.5$ \citep{mahony16,chyzy18,an24}. Assuming such a spectral index yields 0.3\,dex higher luminosities at 1.4\,GHz but no significant difference at 144\,MHz.  For the four diffuse sources, we adopt the galaxy group redshift \citep[i.e., $z=0.2$;][]{wen24} and assume a radio spectral index $\alpha=1.7$, based on the value estimated in Sect.~\ref{sec:radio_spectrum}. Assuming the spectral index derived for the integrated emission, $\alpha=1.2\pm0.2$, the 144\,MHz luminosities remain unchanged, and the 1.4\,GHz luminosities increase by 0.5\,dex. The luminosities at 144\,MHz and 1.4\,GHz of all radio sources are listed in Table~\ref{tab:radio_lum}.  We also report the total luminosities of the four diffuse sources (HSS, envelope, and blob) combined. 
\begin{table}
\caption{\label{tab:radio_lum}Radio luminosities.}
\centering
\renewcommand{\arraystretch}{1.2}
\begin{tabular}{cc lcc}
\hline\hline
 LoTSS  &     ID    & $z_{\rm phot}$        & log(P$_{\rm 144\,MHz}$) & log(P$_{\rm 1.4\,GHz}$) \\
 ID     &           &                       &   (W Hz$^{-1}$)   & (W Hz$^{-1}$) \\
\hline
   85256 & envelope & 0.20\tablefootmark{a} &  24.65$\pm$0.07   &   22.97$\pm$0.52 \\
   16903 & 150      & 0.20                  &  23.35$\pm$0.04   &   22.56$\pm$0.33 \\
   16904 & 58       & 0.08                  &  23.12$\pm$0.03   &   22.33$\pm$0.32 \\
   16905 & 91       & 0.05                  &  22.57$\pm$0.04   &   21.78$\pm$0.33 \\
   16906 & HSS\tablefootmark{a}             & 0.20\tablefootmark{b} &  23.92$\pm$0.08   &   22.24$\pm$0.53 \\
   16907 & 3144     & 0.20\tablefootmark{a} &  23.02$\pm$0.13   &   22.23$\pm$0.42 \\
   16908 & HSS\tablefootmark{a}             & 0.20\tablefootmark{b} &  23.90$\pm$0.09   &   22.22$\pm$0.54 \\
   16909 & east blob & 0.20\tablefootmark{a}&  23.56$\pm$0.10   &   21.90$\pm$0.55 \\
\hline
 Multiple& Diffuse  & 0.20\tablefootmark{a} & 24.81$\pm$0.07 & 23.13$\pm$0.52 \\ 
\hline
\hline
\end{tabular}\\
\tablefoot{The radio luminosities were derived assuming a power-law radio spectrum with spectral index $\alpha=0.8$ for the radio
sources with a galactic counterpart (IDs 150, 58, 91, and 3144), and with $\alpha=1.7$ for the diffuse components (envelope, HSS, and east blob). Assuming a spectral index $\alpha=0.5$ for the galaxies
\citep{an24} yields 1.4\,GHz luminosities that are $\sim$0.3\,dex higher, whereas a spectral index $\alpha=1.2$ for the diffuse sources yields 1.4\,GHz luminosities that are $\sim$0.5\,dex higher. A different spectral index, instead does not change the 144\,MHz luminosities significantly ($<$0.05\,dex). The luminosity errors account for the spectral index uncertainty.
\tablefoottext{a}{Horseshoe-shaped structure.}
\tablefoottext{b}{The redshift is fixed to that of the galaxy group.}}
\end{table}
\section{The nature of J1248$+$4826 through a comparison with other diffuse radio sources}\label{sec:comparison}
To ascertain the true nature of J1248, we compared its observed properties against several established classes of diffuse radio emission. For this analysis we mainly used the following parameters: radio morphology, size, spectrum, and luminosity, host properties, and environment. The results are summarized in Table ~\ref{tab:summary}.

\begin{table*}
\caption{\label{tab:summary}J1248$+$4826 classification.}
\centering
\renewcommand{\arraystretch}{1.2}
\begin{tabular}{ccccccc}
\hline\hline
 Classification             &  Morphology      & Radio       & Radio        & Host       & Environment  & Overall \\
                            &  \& Size         & spectrum    & luminosity   & properties & Halo mass    & interpretation \\
\hline
Classical ORC               & \XSolidBrush & \CheckmarkBold & \CheckmarkBold & \XSolidBrush & \CheckmarkBold & Disfavored  \\
Circular diffuse source     & \CheckmarkBold &  \CheckmarkBold \& \XSolidBrush & \CheckmarkBold & \XSolidBrush & \CheckmarkBold & Disfavored \\
Radio halo/mini-halo        & \XSolidBrush &  \XSolidBrush & \CheckmarkBold \& \XSolidBrush  & \XSolidBrush & \XSolidBrush & Strongly disfavored \\
Stripped ISM                & \XSolidBrush & \XSolidBrush & \CheckmarkBold & \XSolidBrush & \CheckmarkBold & Disfavored  \\
Active radio galaxy & \XSolidBrush & \CheckmarkBold & \CheckmarkBold & \XSolidBrush & \nodata  & Strongly disfavored  \\
(Re-energized) remnant lobe & \CheckmarkBold \& \XSolidBrush & \CheckmarkBold & \CheckmarkBold & \CheckmarkBold & \CheckmarkBold & Favored \\
\hline
\hline
\end{tabular}\\
\end{table*}
\subsection{Is J1248$+$4826 an odd radio circle ?}
The apparent ring-like morphology of J1248 (see Fig.~\ref{fig:radio_on_multilambda}) initially drew comparisons with the so-called ORCs \citep{norris21_orc_discovery,norris21_review,koribalski21}. However, the ORC population has become increasingly heterogeneous as new sources have been reported. A compilation of all sources currently designated as ORCs in the literature, together with their host galaxy properties and radio characteristics, is presented in Table~\ref{tab:lit_orcs}. In Appendix~\ref{app:orcs}, we review the observed diversity of these sources and discuss whether all satisfy the canonical definition proposed by \citet{norris25}. This context provides the framework for assessing whether J1248 should be classified as an ORC based on both the properties of the known population and the formal observational definition. 

Assuming that the residual diffuse emission in J1248 is associated with the WHJ124851.9+482616 galaxy group at $z=0.2$, its redshift is consistent with the known ORC population, which spans $z = 0.02 - 0.86$ with a median of 0.30. The apparent ring-like structure in J1248 has a radius of roughly 30\,kpc, which is notably smaller than the radii of literature ORCs (44\,kpc to 365\,kpc). However, the surrounding diffuse envelope extends to a radius of $\sim100$ kpc, aligning well with the typical extent of ORCs. The spectral indexes derived in Sect.~\ref{sec:radio_spectrum} for the integrated source ($\alpha_{\rm tot}=1.2\pm0.2$) and the diffuse emission ($\alpha_{\rm diffuse}=1.7\pm0.2$) are both consistent with the steep spectra generally observed in ORCs \citep[i.e., $\alpha{\sim}0.8-1.6$;][]{norris22,dolag23,lochner23}. Furthermore, using the procedure described in Sect.~\ref{sec:radio_lums}, we estimated the 150\,MHz luminosity of the diffuse component to be log\,(P$_{\rm 150\,MHz}$/[W\,Hz$^{-1}] = 24.78\pm0.02$, well within the observed ORC range (log\,(P$_{\rm 150 MHz}$/[W\,Hz$^{-1}$]) = 23.3--26.8).\null

\begin{figure}[!ht]
\centering
\includegraphics[width=0.5\textwidth]{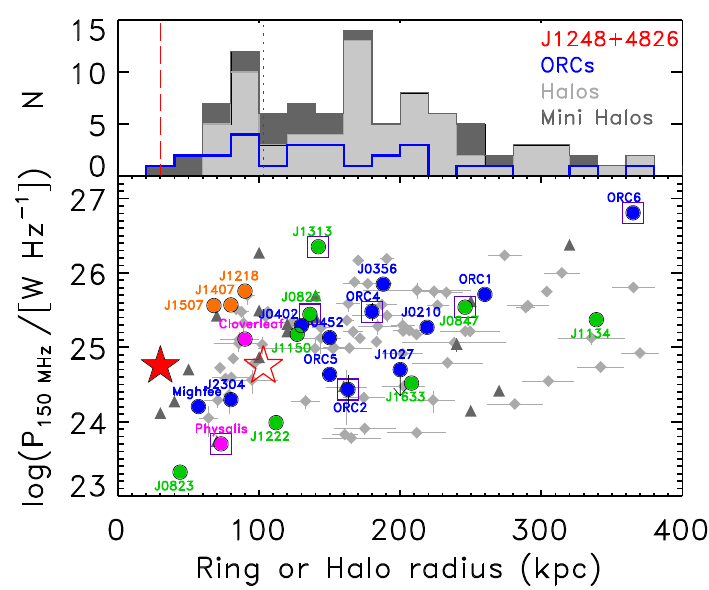}
  \caption{Radio powers and sizes of ORCs, radio halos and mini-halos. \textit{Main panel: } Radio power at 150\,MHz as a function of the radius for J1248 (filled red star) and for several ORCs from the literature (filled circles): edge-brightened ORCs (blue), shell-like ORCs in groups (magenta), diffuse circles (orange), and LOFAR ORCs (green; see Table~\ref{tab:lit_orcs}). The radius of the full extent of J1248 is shown with an empty red star.  The purple squares represent the double-headed ORCs. The ORCs are labeled by name. The filled gray diamonds represent a sample of radio halos associated with \planck-selected galaxy clusters found in the LoTSS survey \citep{botteon22}. The filled dark-gray triangles refer to mini-halos from the collection in \citet{giacintucci14}. \textit{Top panel:} Distribution of radii for ORCs (solid blue line), radio halos (light-gray-filled histogram), and mini-halos (dark-gray-filled histogram). The radius of the HSS and of the whole diffuse emission of J1248 are marked by a vertical dashed and dotted red line, respectively. }
\label{fig:Prad_size}
\end{figure}

A quantitative comparison is presented in Fig.~\ref{fig:Prad_size} where we show the radio luminosity as a function of size for J1248 and for other ORCs (see Table~\ref{tab:lit_orcs}).  Overall, larger structures are more luminous, although, at a fixed size, the luminosity spans roughly two orders of magnitude. J1248 stands out in this P$_{\rm 150\,MHz}$ -- radius diagram because of its small size. Its luminosity is consistent with the ORCs median luminosity (i.e., log(P$_{\rm 150\,MHz}$/[W\,Hz$^{-1}$])=$25.3\pm0.6$), but its radius, 30\,kpc or 9\arcsec, is the smallest ever reported. The radii of ORCs in the literature range from 44\,kpc to 365\,kpc (equivalent to 0.3\arcmin--3.4\arcmin), and the median radius is 150\,kpc, five times larger than the radius of J1248.  However, if we consider its full extent ($\sim$100\,kpc radius), J1248 lies well within the ORC distribution.

\begin{figure}[!ht]
\centering
\includegraphics[width=0.5\textwidth]{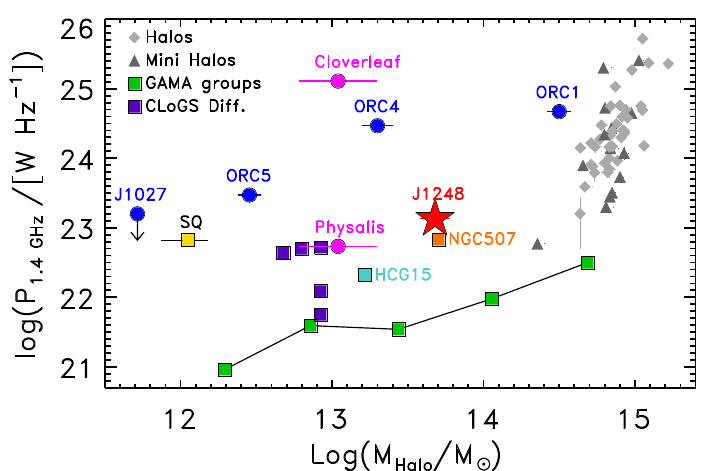}
  \caption{Radio power at 1.4\,GHz and halo mass for J1248 (filled red star), a few ORCs (filled blue circles: edge-brightened ORCs; filled magenta circles: shell-like ORCs in groups), cluster radio halos \citep[light-gray diamonds;][]{botteon22,vanweeren21,cassano13}, mini-halos \citep[dark-gray triangles;][]{giacintucci14}, diffuse emission from CLoGS \citep[purple squares;][]{kolokythas19}, the remnant plasma in \object{HCG\,15} \citep[turquoise square;][]{riseley25}, diffuse components in the \object{NGC\,507} galaxy group \citep[orange square;][]{brienza22} and SQ \citep[yellow square;][]{moles97}, and average values from 400 GAMA galaxy groups \citep[green squares and black line;][]{wagh26}.} 
\label{fig:Prad_Mhalo}
\end{figure}

Another common trait J1248 shares with many ORCs is its environment. Several ORCs reside in galaxy groups with intermediate mass halos (M$_{\rm vir}\sim10^{12-13}$\,\msun; see Table~\ref{tab:lit_orcs} and Fig.~\ref{fig:Prad_Mhalo}). The total halo mass of the $z=0.2$ group in the J1248 field, estimated from the total stellar mass of the member galaxies is $\sim6\times 10^{13}$\,\msun\ \citep{wen24}. We derived a slightly lower value, M$_{\rm halo}\simeq(3.8-4.9)\times 10^{13}$\,\msun\ from the mass of the most massive group member, ID 145, assuming the stellar-to-halo mass relation (SHMR) at $z=0.2$ from \citet{girelli20}. Although both estimates suffer from substantial ($\lesssim$35\%) uncertainties inherent to the employed method, the halo mass of J1248 is consistent with the range of halo masses observed in other ORCs (i.e., (0.05--40)$\times10^{13}$\msun). \null

Morphologically, however, J1248 diverges from most ORCs. It does not display the nested shells of Cloverleaf and Physalis, and its apparent ring appears to be a HSS surrounded by diffuse emission. 
Finally, J1248 does not host a central ETG, with the closest candidate (ID 145) being 16\arcsec\ away from the center (see Fig.~\ref{fig:radio_residuals}). Interestingly, J1248 shares the closest resemblance with ORC2, displaying an extended diffuse envelope and an offset host galaxy, but this system is now understood to be the lobe of a restarted radio galaxy.\null

In summary, J1248 shares with ORCs a steep spectrum, high low-frequency luminosity, and a group environment. However, its lack of a distinct limb-brightened ring, the presence of an extended surrounding envelope, and of an offset passive galaxy are atypical of classical ORCs. Given these features, it is perhaps not surprising that J1248 was not included in the sample of 27 ORCs and candidates recently identified by \citet{degasperin26} in LoTSS DR3, despite meeting all of their initial selection criteria (total-to-peak flux ratio $>10$, angular size $>40^{\prime\prime}$, low local noise, and high latitude). We therefore disfavor a standard ORC interpretation for J1248.\null

\subsection{Is J1248$+$4826 a circular diffuse source ?}

Circular diffuse sources represent another recently discovered class of puzzling diffuse radio sources that are sometimes designated as ORCs \citep{kumari24,kumari25,gupta25,kumari24_horseshoe}.  In these sources, the host is usually a bright radio source slightly offset from the geometric center, and the radio surface brightness distribution is characterized by a bright central peak that gradually fades outward. Overall, their partially filled interiors and fragmented edges indicate a complex internal structure that is atypical in classical ORCs. \citet{kumari24_horseshoe}  suggested that they may represent a new class of luminous, extended diffuse sources. The best studied examples of this class are listed in the fourth group of Table~\ref{tab:lit_orcs}. 

For our analysis, one system in particular, J1407$+$0453 \citep{kumari24_horseshoe}, is intriguing due to its morphological resemblance to J1248.  This source features a diffuse circular component with a total extent of 65\arcsec, a horseshoe-shaped structure, 10\arcsec\ in diameter, at its center, and a host galaxy situated directly on the limb of the horseshoe structure. Nonetheless, there are two significant physical discrepancies with J2148.  First, J1407$+$0453 exhibits a relatively flat radio spectrum ($\alpha=0.67$), contrasting sharply with the steep spectrum observed in J1248. Second, this new class of circular diffuse sources appears systematically more luminous than other diffuse radio structures of comparable size, including J1248 (see orange symbols in Fig.~\ref{fig:Prad_size}). The scarce number of well-studied sources belonging to this class and the lack of a clear definition and distinct properties make it difficult to place any source within this category. However, the lack of a radio bright host in J1248 leads us to disfavor this classification. 

\subsection{Is J1248$+$4826 a radio halo or mini-halo ?}
The extended diffuse radio emission of J1248 and its connection with a group is reminiscent of similar structures found in the centers of galaxy clusters: radio halos and mini-halos \citep{vanweeren19}. Radio halos are large, diffuse radio emission found in the central regions of galaxy clusters, and are detected in about 30\% of galaxy clusters \citep{botteon22}. They are thought to be produced by reacceleration of relativistic electrons due to shock waves triggered by merging clusters. Mini-halos, conversely, are smaller ($\sim100-500$ kpc) and typically surround a central, powerful radio galaxy in relaxed clusters.\null

Interestingly, J1248, radio halos, and mini-halos all share similar 150\,MHz radio luminosities. As illustrated in Fig.~\ref{fig:Prad_size}, the radio luminosity of J1248 is within the range of those derived for \planck-selected cluster halos and mini-halos \citep{giacintucci14,botteon22}. Its extent is consistent with the sizes of the smallest halos and mini-halos.  However, critical differences emerge in their morphologies and host environments.  While radio halos and mini-halos typically exhibit a surface brightness profile that peaks strongly at the center, the diffuse emission of J1248 lacks a single central peak (see Fig.~\ref{fig:orc_profile}).  Furthermore, as shown in Fig.~\ref{fig:Prad_Mhalo}, radio halos and mini-halos are almost exclusively hosted by massive dark matter halos, whereas the estimated halo mass for the J1248 group is approximately an order of magnitude lower. J1248 is also less luminous at 1.4\,GHz. This is due to its steeper radio spectrum; indeed, the average spectral indexes of halos and mini-halos are, respectively, $\alpha\simeq1.3$ and 1. Given the significant discrepancies in halo mass and surface brightness profile, we conclude that J1248 cannot be classified as a radio halo or mini-halo.\null

\begin{figure}[!ht]
\centering
\includegraphics[width=0.48\textwidth]{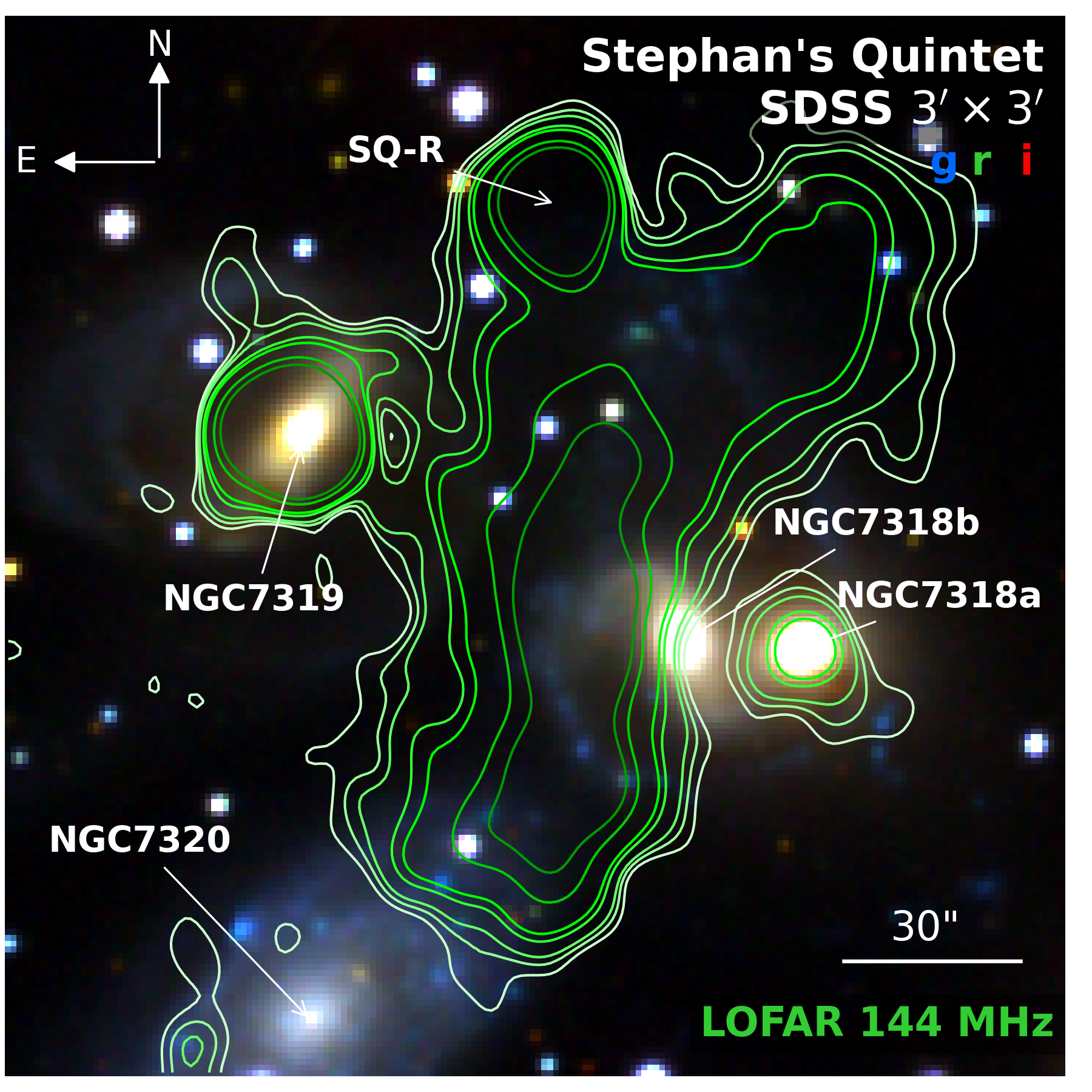}
 \caption{SDSS image of the SQ compact group at $z=0.0215$ (blue: g band; green: r band; and red: i band). The group galaxies (NGC\,7318a, NFC\,7318b, and NGC\,7319), a foreground galaxy (NGC\,7320) and a radio source (SQ-R) are marked. The green contours represent the LOFAR 144 MHz emission (at 3, 4, 5, 8, 10, 20, and 30$\sigma$, with $\sigma$=0.1\,mJy\,beam$^{-1}$).  }
\label{fig:SQ}
\end{figure}
\subsection{Is J1248$+$4826 the result of galaxy interactions ?}\label{sec:interaction}

The foreground radio-bright galaxies in J1248, IDs 58, and 91, lie at similar redshifts \citep[$z_{\rm phot} = 0.07 \pm 0.05$ and $0.05 \pm 0.01$, respectively;][]{wen24} and exhibit irregular morphological features indicating that they might be interacting. A prime example of a galaxy interaction that has generated a diffuse radio source on large scales is represented by the \object{Stephan Quintet} (SQ) compact group. The SQ comprises a group of galaxies at $z \simeq 0.0215$ \citep{hickson92} whose interaction has stripped a large fraction of the galaxies' ISM into the IGM \citep{sulentic01}. This mechanism lies at the origin of a giant, multiphase filament of shocked gas approximately 45\,kpc in length \citep{arnaudova24} that emits across radio, X-ray bands, and in \Ha.  The LOFAR 144\,MHz image of the SQ against its optical image is shown in the left panel of Fig.~\ref{fig:SQ}. The similar appearance of J1248 and the SQ motivates us to explore whether the same mechanism might explain the diffuse radio emission in J1248. To test this hypothesis, we compared the radio properties of J1248 and of the SQ assuming that galaxies IDs 58, 91 and the diffuse radio component are at $z \sim 0.05$.  

Based on the LoTSS DR3 deblended catalog, approximately 63\% of the total flux density at 144\,MHz in the SQ region comes from the diffuse filament. In J1248, this fraction is slightly higher, with $\sim 73\%$ of the total emission coming from the diffuse component. If placed at $z \sim 0.05$, the 144\,MHz luminosity of J1248's diffuse component would be $\log({\rm P}_{\text{144 MHz}} / [\text{W Hz}^{-1}]) = 23.48 \pm 0.07$, consistent with the SQ filament luminosity ($\log({\rm P}^{\rm SQ}_{\text{144 MHz}} / [\text{W Hz}^{-1}]) = 23.684 \pm 0.001$).

However, SQ and J1248 differ significantly. First, J1248 exhibits a much steeper radio spectrum ($\alpha = 1.7 \pm 0.1$) than the SQ filament ($\alpha \sim 0.87 \pm 0.16$). Second, their spatial configurations contrast sharply. While SQ's $1.9^{\prime}$ filament perpendicularly bisects its aligned galaxies (which are separated by $1.6^{\prime}$), the diffuse emission in J1248 sits entirely to the east of galaxies ID 58 and 91, spanning an area three times larger than their $0.5^{\prime}$ separation. Finally, unlike SQ, J1248 lacks any visible stellar trails connecting the interacting galaxies. Given these discrepancies, it is unlikely that J1248 is analogous to the SQ. Obtaining spectroscopic redshifts for the pair, and X-ray or \Ha\ observations could definitively test this scenario by confirming the proximity and by mapping the precise location of any shocked gas, as done successfully for SQ \citep{sulentic01,osullivan09,arnaudova24}.

\subsection{Is J1248$+$4826 the lobe of an active radio galaxy ?}\label{sec:radio_galaxy}

Depending on geometries and viewing angles, radio jets and lobes can appear as diffuse radio sources with some structures \citep[see, e.g.,][]{hota25}. We therefore examined the possibility that the diffuse emission in the field of J1248 might be due to a jet or lobe powered by a currently active radio galaxy.\null

Only three galaxies in the field, IDs 58, 91, and 150, are detected in the radio. All three are classified star-forming galaxies and lack clear evidence of AGN activity, though a deeply obscured or intrinsically weak AGN cannot be ruled out with the current data. Since IDs 58 and 91 are low mass galaxies (M$_{\rm star}<10^{10}$\,\msun) and only massive galaxies typically host powerful radio emission \citep{uchiyama22}, it is unlikely that they could drive a powerful jet and deposit enough plasma to generate the diffuse radio emission in J1248. However, as ID 150 is a relatively massive (M$_{\rm star}=(7\pm1)\times 10^{10}$\,\msun) galaxy, we evaluated the possibility that it is at the origin of the diffuse radio component. The morphology of J1248 could theoretically be interpreted as a single radio lobe originating from ID 150, situated approximately 50\arcsec\ ($\sim 170$ kpc) from the center of the diffuse emission.  This physical connection is tentatively supported by the presence of a faint radio bridge (LoTSS source ID 16907) linking this galaxy to the diffuse structure.\null

However, there are three issues with this interpretation: (i) the absence of a counter lobe,  (ii) the late type classification of ID 150, and (iii) the lack of a radio flux in excess compared with that derived from the SFR. First, the absence of a counter-lobe is not unprecedented, as similar asymmetric structures have been observed in other systems \citep[e.g., the brightest group galaxy in SDSSTG28674;][]{santra26}.  The presence of a single lobe could result from gas motions that compress and re-energize only one side \citep[e.g.,][]{randriamanakoto20,shulevski24},  the counter-lobe aging out of the observable band after
expanding into a lower density region \citep[e.g.,][]{giacintucci25}, the host galaxy drifting away from the center \citep[e.g.,][]{shulevski24}, and/or having the two lobes superimposed along the line of sight. Second, the intense star-forming activity, the lack of AGN signatures, and the disk profile\footnote{ID 150 is listed in the DESI catalog under the "Sersic" morphological type based on the Tractor classification \citep{hogg13}. This implies a disk morphology as expected for a late-type galaxy.} of ID 150 contrast with the standard radio galaxy properties. Nevertheless, some rare cases of radio galaxies hosted by star-forming disks exist in the literature \citep{singh15,dabhade20,bagchi25,tate26}. Finally, the radio emission from ID 150 is fully consistent with its SFR. This implies that there is no AGN-driven radio activity in the core of this galaxy, contrary to what would be expected for a radio galaxy. The first two factors would make ID 150 an unusual radio galaxy, but the latter rules out this interpretation.\null

There is, however, another galaxy in the field that exhibits properties typical of radio galaxies, such as being massive and passive, yet it is not detected by LOFAR. The galaxy ID 145 is the most massive galaxy in the group, and both its SED (see Fig.~\ref{fig:seds}) and the optical SDSS spectrum \citep{sdss_dr6} indicate that it is a passive galaxy (red in the visible, strong Balmer break and absorption features, no emission lines, and no mid-IR emission). Although not individually detected at 144 MHz, ID 145 is part of the HSS and of source LoTSS ID 16908. In addition, it is coincident with a relatively bright blob in the LoLLS image at 54\,MHz. Based on a visual comparison, ID 145 seems to have a steeper radio spectrum than the HSS and the surrounding envelope. Its superposition to the diffuse emission and its proximity (16\arcsec) to the center of the HSS hint to the possibility that J1248 might be a lobe ejected by ID 145 nearly along the line of sight, analogous to the precessing jet model proposed for ORCs by \citet{nolting23}.  However, according to this model the base of the jet should be radio bright, the spectrum flat, and the host blazar-like, contrary to the apparently steeper spectrum and the relatively faint radio emission of ID 145. Furthermore, there are no signs of AGN activity in this galaxy from the SED, the spectrum, or the radio luminosity.  There is only one source in the group with signs of AGN activity from the SED, ID 214, but it is not a radio emitter, it does not overlap with the diffuse radio emission, and it is neither luminous nor massive. Consequently, we conclude it is highly unlikely that J1248 is powered by ongoing AGN activity.\null
\subsection{Is J1248$+$4826 a remnant lobe ?}\label{sec:fossil}
Alternatively, J1248$+$4826 may represent the remnant lobe of an extinct or dying AGN. Once the central engine shuts off, the remnant radio lobe expands into the surrounding medium, undergoes radiative and adiabatic losses, becomes more diffuse and progressively fades over time \citep[e.g.,][]{cordey87,brienza16,brienza22}. In the absence of a continued energy input, the electron population cools and the emission declines on timescales of order $\sim$10--100\,Myr \citep{enblin03}. The radio spectrum steepens significantly ($\alpha\sim1.2-2$), making these structures primarily detectable at sub-gigahertz frequencies. Such systems typically lack a visible radio core or compact radio structures, exhibiting amorphous or double-lobed morphologies with low surface brightness \citep[$\sim10-30$\,mJy\,arcmin$^{-2}$;][]{brienza17} and spectral flattening at the lowest frequencies \citep[$\Delta\alpha>0.5$;][]{murgia11}. The progenitors of these structures are typically massive galaxies with, on average, a stellar mass of $\sim2\times10^{11}$\,\msun\ \citep{jurlin20,jurlin21}.\null

We evaluated whether the properties of J1248 align with this dying AGN scenario. The most plausible progenitor of the fossil plasma is the massive group member ID 145, given its close proximity to the center of the diffuse emission ($\sim$16\arcsec\ or 55\,kpc), its high stellar mass (M$_{\rm star}=(2.6\pm0.1)\times 10^{11}$\,\msun) and its passive status.  Its optical spectrum, characterized by prominent absorption features and a lack of emission lines, is typical of ETGs \citep{kennicutt12} and indicates that the central engine is inactive. If the diffuse emission is a remnant lobe of ID 145, the absence of a visible counter-lobe could stem from the geometric or dynamical effects discussed in the previous section. Furthermore, consistent with a dying AGN, the galaxy's core radio emission is faint and its spectrum steep. While the average surface brightness in the LOFAR residual image (i.e., $\sim$43\,mJy\,arcmin$^{-2}$) is somewhat higher than typically found in remnant lobes, the low peak-to-total flux ratios of the diffuse components (i.e., $\lesssim$0.2; see Table~\ref{tab:lofar_data}) are well within the expected range for a dying AGN \citep{degasperin18}. We therefore conclude that J1248 might be an old remnant lobe of the massive galaxy ID 145.

\subsection{Final remarks on the nature of J1248$+$4826}

Our comparative analysis is summarized in Table~\ref{tab:summary}. On the one hand, the steep radio spectrum, high luminosity, and group environment of J1248 align closely with established populations, including ORCs, circular diffuse sources, and remnant lobes. However, its specific structural features tell a different story: the unique morphology of a highly compact HSS embedded within an extended diffuse envelope, combined with an offset passive host galaxy, differs from the standard expectations of any single category. Nevertheless, its numerous properties in common with fossil plasma, coupled with the absence of an active radio galaxy in the field, strongly favor a remnant lobe interpretation, with the inactive massive group member ID 145 as its putative progenitor.

\section{J1248$+$4826 formation scenarios}\label{sec:discussion}
The comparison above favors a remnant lobe origin for J1248. Since no radio galaxy in the field is currently sufficiently active to power this structure (see Sects.~\ref{sec:radio_galaxy}), the persistence of its bright diffuse radio emission implies either the recent cessation of activity (a dying radio galaxy) or the re-energization by external processes, such as turbulence and shock waves in the surrounding medium. Below we discuss whether these mechanisms can explain the properties of J1248.
\subsection{A dying radio galaxy}

In Sect.~\ref{sec:fossil} we identified the massive, currently quiescent galaxy ID 145 as the putative progenitor. Here, we specifically ask whether the recent cessation of activity of ID 145 can solely account for the observed properties of J1248.  The faint radio emission at 54\,MHz of ID 145 hints at past activity, but its spectrum is visibly steeper than the extended diffuse envelope. This spectral difference implies that the electron populations in the core are older than in the remnant lobe \citep{hardcastle18}. The flatter spectrum of the lobe with respect to its progenitor cannot be reconciled even in cases of physical confinement by the thermal pressure of a dense external medium. While environmental confinement would slow the adiabatic expansion rate and extend the lobe's observable lifetime \citep[][but see also \citealt{jurlin21}]{murgia11}, it would ultimately cause the remnant lobe's spectrum to steepen even further over time due to radiative cooling \citep{blandford19}. Therefore, the flatter spectrum and enhanced HSS dictate that the plasma must have
undergone an additional reacceleration event after its initial deposition.

\subsection{Galaxy mergers in a group as re-energizing mechanism}\label{sec:mechanisms}

Given the need for an external re-energization mechanism, the host environment becomes crucial. The recent discovery of several diffuse radio sources in galaxy groups \citep[e.g.,][]{brienza22,koribalski24_physalis,koribalski24,bulbul24,riseley25,degasperin26} strongly points to group dynamics as a primary driver.

Galaxy groups are characterized by intermediate halo masses (M$_{\rm halo}=10^{12.5}-10^{14}$\,\msun), where the low velocity dispersions ($\sim150-400$\,\kms) and the close proximity of member galaxies favor mergers and tidal interactions \citep{kolokythas18}. Hydrodynamic simulations have shown that merging group members can generate turbulence and moderate shock waves within the IGrM \citep{dolag23,ivleva26}.  The lobe can be considered as an underdense bubble of plasma in the IGrM. In the presence of a shock wave, the bubble is compressed, differential motion may develop, and vorticity might be deposited at the interface, causing the formation of synchrotron emitting structures around a central depression \citep{wang26}. This re-energized plasma can produce irregular extended radio structures, full or patchy rings, shells, and complex substructures such as the ring or the HSS observed in J1248 \citep{dolag23,Lin_Yang24}.\footnote{The complex morphologies reproduced by this scenario are visible at http://www.magneticum.org/complements.html\#Compass.} Because these structures are produced by external shocks, they do not require a central host galaxy, their morphology is dictated instead by the initial jet orientation and the geometry of the shock front.

We can estimate the required shock strength using standard diffusive shock acceleration \citep[DSA;][]{blandford87}, where the Mach number ($\mathcal{M}$) is derived from the radio spectral index ($\alpha$) as $\alpha =(\mathcal{M}^2+1)/(\mathcal{M}^2-1)$ \citep[for the specific assumptions behind this relation, see][]{wittor21}. The total spectral index of the J1248 region ($\alpha = 1.2$) and the diffuse component ($\alpha = 1.7$) yield Mach numbers of $\mathcal{M}_{\text{total}} = 3.3$ and $\mathcal{M}_{\text{diffuse}} = 2.0$, respectively.  These values align perfectly with the moderate shocks predicted by the merging group members model.

However, the observed radio power of J1248 (i.e., log(P$_{\rm 150 MHz}/{\rm[W\,Hz^{-1}])}=24.78\pm0.02$) is several orders of magnitude higher than the value predicted by the simulations \citep[i.e., $\sim10^{21}$\,W\,Hz$^{-1}$ at 150\,MHz;][]{dolag23}. This discrepancy is easily explained by the physical nature of the shocked material. In \citet{dolag23}, the shock sweeps up the standard thermal pool of electrons in the circumgalactic medium (CGM). In contrast, if J1248 represents a remnant radio lobe, the passing shock is interacting with fossil plasma that already contains a rich population of highly energetic electrons. In this scenario, even a moderate merger shock can reaccelerate these relativistic electrons with very high efficiency, easily achieving the observed high radio luminosity of J1248. While this scenario is plausible, definitively confirming it first requires identifying merging group galaxies in the field, and subsequently running specific simulations. Such models require currently unavailable constraints, including the precise IGrM pressure profile, local magnetic field strengths, and a spatially resolved spectral-age distribution of the plasma.

\subsection{Comparison with diffuse radio emission in galaxy groups}

To understand the high luminosity of J1248, we examined it with respect to broader group scale trends. A recent wide-field survey using ASKAP data reports that diffuse radio emission in groups is relatively rare, appearing in $\sim 11\%$ of sampled systems, with typical 1.4\,GHz radio powers spanning $10^{19}$ to $10^{24}$ W Hz$^{-1}$ \citep[median $\sim 3.5 \times 10^{21}$ W Hz$^{-1}$;][see also \citealt{nikiel19}, and \citealt{kolokythas19}]{wagh26}. As illustrated in Fig.~\ref{fig:Prad_Mhalo}, the average diffuse emission in groups generally sits above the simple mass-scaled extrapolations of cluster radio halos.  However, J1248, alongside ORCs and a handful of well studied group sources (NGC\,507, \citealt{brienza22}; HCG\,15, \citealt{riseley25}; SQ, \citealt{moles97}; and the most luminous examples from the complete local-volume groups sample (CLoGs), \citealt{kolokythas19}) stands out as 1–-3\,dex more luminous than the groups average values. This stark contrast demonstrates that halo mass is not the primary parameter governing the radio power of diffuse synchrotron sources in low-mass systems. An additional ingredient, such as a particularly rich seed population of relativistic electrons, unexpectedly strong local magnetic fields, or a particularly efficient re-energization event is required. These types of sources thereby provide a direct window into the recent history of AGN activity, group dynamics, and their long-term coupling.

\subsection{Testing the proposed formation scenario}

While our multiwavelength analysis strongly favors the interpretation of J1248 as a remnant radio lobe reaccelerated by group dynamics, further observational evidence would be necessary to confirm and refine this formation scenario. High-resolution, multifrequency radio observations (spanning from sub-gigahertz LOFAR up to 3–-6\,GHz with the VLA) are essential. Spatially resolved spectral index maps will allow us to trace age gradients across the diffuse envelope, identify the precise sites of energy injection as a flattening of the spectrum at the leading edge of the shock, followed by a sharp steepening in the downstream (post-shock) region, and search for the high-frequency spectral curvature characteristic of aging fossil plasma. If the plasma is re-energized by turbulence, the spectral index map will appear patchy. Measuring the polarization of the diffuse emission will provide a robust test for the merger-shock model. When a shock passes through a remnant lobe, it compresses the plasma and the magnetic fields.  This creates a highly ordered magnetic field aligned parallel to the shock front. In the case of a fading lobe without shock compression, the magnetic field would be relaxed, resulting in low polarization.  Similarly, turbulent reacceleration isotropizes the magnetic field, which heavily depolarizes the emission.

Deep X-ray observations are crucial for characterizing the IGrM. Mapping the thermal pressure profile will determine if the external environment is dense enough to confine the remnant lobe and prolong its observable lifetime. Furthermore, high-sensitivity X-ray data could enable the direct detection of the proposed merger-driven shocks via surface brightness or temperature jumps. A comprehensive spectroscopic survey of the galaxies in the J1248 field is also needed to confirm group membership, accurately measure velocity dispersions, and identify the specific interacting or merging members responsible for driving the turbulence and shocks into the IGrM.

\section{Summary and conclusions}\label{sec:conclusions}
We report the discovery and multiwavelength analysis of J1248$+$4826, a complex diffuse radio source identified in LOFAR 144 MHz observations. After removing the radio emission from resolved radio galaxies, the residual emission reveals a compact ($\sim30$ kpc radius) horseshoe-shaped structure embedded in an extended ($\sim200$ kpc) diffuse envelope. This residual component contains approximately 73\% of the total radio flux and has a steep radio spectrum ($\alpha=1.7\pm0.2$). Our interpretation is that it is a remnant radio lobe. Its most plausible progenitor is a massive and passive galaxy, a member of a galaxy group at $z=0.2$ and located approximately 55\,kpc from the radio structure center. The morphology and spectrum of this diffuse radio source suggest that it is undergoing re-energization, although the available data do not yet specify the mechanism. Moderate ($\mathcal{M} \sim 2-3$) shocks associated with the group dynamics may provide a plausible explanation, though additional data and simulations are required to test this.
 
The initial 144\,MHz image of J1248$+$4826 displays several traits that resemble ORCs and circular diffuse sources, including an apparent ring-like structure, comparable radio luminosities, and a steep spectral shape. However, J1248$+$4826 ultimately diverges from these classes. It breaks from the strict ORC definition due to its offset potential host galaxy, extended diffuse envelope, and the distinctly horseshoe-shaped morphology revealed after subtracting the galaxy emission. While these morphological traits are frequently seen in circular diffuse sources, J1248$+$4826 differs from that population as well by exhibiting a steeper radio spectrum, a lower overall luminosity, and the lack of a radio bright host. This study illustrates the challenge of establishing the physical origins of diffuse radio sources and the need for clear definitions of source types in view of an increasing number of diffuse radio sources waiting to be revealed.

To definitively constrain the origin of J1248$+$4826 further observations are needed. High-resolution and multifrequency radio observations will be important for resolving the spectral properties and internal structure. Spatially resolved spectral index maps are needed to trace age gradients and identify precise sites of energy injection, while polarization measurements would robustly test the shock model by revealing the underlying magnetic field structure. Finally, sensitive X-ray observations will be crucial for mapping the pressure profile of the IGrM and for directly detecting the proposed shocks. Ultimately, expanding the sample of similarly complex diffuse structures with high resolution, multiwavelength follow-up observations will be essential to untangle their diverse origins, establish their formation, and decipher their evolutionary pathways. Discovering these diffuse radio sources is akin to briefly turning the lights on in a dark room, suddenly revealing the hidden history and impact of its occupants. In much the same way, these structures reveal the energetic past of currently inactive galaxies and map the large-scale dynamical evolution of their environments.

%%%%%%%%%%%%%%%%%%%%%%%%%%%%%%%%%%%%%%%%%%%%%%%%%%%%%%%%%%%%%%
\begin{acknowledgements}
The authors thank the referee for an insightful report and useful suggestions that significantly improved the analysis.
M.P. warmly thanks Dr. Fabio Gastaldello and Dr. Konstantinos Kolokythas 
for constructive and useful suggestions.
M.P. kindly thanks Dr. Zhong Lue Wen for providing magnitudes and photometric
redshifts from the DESI DR9 dataset.
%INAF
M.P. acknowledges financial support from INAF mini-grant 2023 "1.05.23.04.01".

%LOFAR
LOFAR data products were provided by the LOFAR Surveys Key Science project (LSKSP; \url{https://lofar-surveys.org/}) and were derived from observations with the International LOFAR Telescope (ILT). LOFAR \citep{vanhaarlem13} is the Low Frequency Array designed and constructed by ASTRON. It has observing, data processing, and data storage facilities in several countries, which are owned by various parties (each with their own funding sources), and which are collectively operated by the ILT foundation under a joint scientific policy. The efforts of the LSKSP have benefited from funding from the European Research Council, NOVA, NWO, CNRS-INSU, the SURF Co-operative, the UK Science and Technology Funding Council and the Jülich Supercomputing Centre.

%VLA
The National Radio Astronomy Observatory and Green Bank Observatory are facilities of the U.S. National Science Foundation operated under cooperative agreement by Associated Universities, Inc.

%WENSS
WENSS is a joint project of the NFRA and Leiden Observatory.

%DESI
The Legacy Surveys consist of three individual and complementary projects:
the Dark Energy Camera Legacy Survey (DECaLS; Proposal ID \#2014B-0404; PIs:
David Schlegel and Arjun Dey), the Beijing-Arizona Sky Survey (BASS; NOAO
Prop.  ID \#2015A-0801; PIs: Zhou Xu and Xiaohui Fan), and the Mayall z-band
Legacy Survey (MzLS; Prop.  ID \#2016A-0453; PI: Arjun Dey).  DECaLS, BASS
and MzLS together include data obtained, respectively, at the Blanco
telescope, Cerro Tololo Inter-American Observatory, NSF’s NOIRLab; the Bok
telescope, Steward Observatory, University of Arizona; and the Mayall
telescope, Kitt Peak National Observatory, NOIRLab.  Pipeline processing and
analyses of the data were supported by NOIRLab and the Lawrence Berkeley
National Laboratory (LBNL).  The Legacy Surveys project is honored to be
permitted to conduct astronomical research on Iolkam Du’ag (Kitt Peak), a
mountain with particular significance to the Tohono O’odham Nation.

NOIRLab is operated by the Association of Universities for Research in
Astronomy (AURA) under a cooperative agreement with the National Science
Foundation.  LBNL is managed by the Regents of the University of California
under contract to the U.S.  Department of Energy.

BASS is a key project of the Telescope Access Program (TAP), which has been
funded by the National Astronomical Observatories of China, the Chinese
Academy of Sciences (the Strategic Priority Research Program “The Emergence
of Cosmological Structures” Grant \# XDB09000000), and the Special Fund for
Astronomy from the Ministry of Finance.  The BASS is also supported by the
External Cooperation Program of Chinese Academy of Sciences (Grant \#
114A11KYSB20160057), and Chinese National Natural Science Foundation (Grant
\# 12120101003, \# 11433005).

The Mayall z-band Legacy Survey (MzLS; NOAO Prop.  ID no.  2016A-0453; PI:
A.  Dey) uses observations made with the Mosaic-3 camera at the Mayall 4 m
telescope at Kitt Peak National Observatory, National Optical Astronomy
Observatory, which is operated by the Association of Universities for
Research in Astronomy (AURA) under cooperative agreement with the National
Science Foundation.  The authors are honored to be permitted to conduct
astronomical research on Iolkam Du’ag (Kitt Peak), a mountain with
particular significance to the Tohono O’odham.

The Legacy Survey team makes use of data products from the Near-Earth Object
Wide-field Infrared Survey Explorer (NEOWISE), which is a project of the Jet
Propulsion Laboratory/California Institute of Technology.  NEOWISE is funded
by the National Aeronautics and Space Administration.

%Herschel
The \herschel\ spacecraft was designed, built, tested, and launched under a
contract to ESA managed by the \herschel/\planck\ Project team by an industrial
consortium under the overall responsibility of the prime contractor Thales
Alenia Space (Cannes), and including Astrium (Friedrichshafen) responsible
for the payload module and for system testing at spacecraft level, Thales
Alenia Space (Turin) responsible for the service module, and Astrium
(Toulouse) responsible for the telescope, with in excess of a hundred
subcontractors.  SPIRE has been developed by a consortium of institutes led
by Cardiff University (UK) and including Univ.  Lethbridge (Canada); NAOC
(China); CEA, LAM (France); IFSI, Univ.  Padua (Italy); IAC (Spain);
Stockholm Observatory (Sweden); Imperial College London, RAL, UCL-MSSL,
UKATC, Univ.  Sussex (UK); and Caltech, JPL, NHSC, Univ.  Colorado (USA). 
This development has been supported by national funding agencies: CSA
(Canada); NAOC (China); CEA, CNES, CNRS (France); ASI (Italy); MCINN
(Spain); SNSB (Sweden); STFC, UKSA (UK); and NASA (USA). The \herschel\ data
are from the DDT program 1305. 

%Spitzer 
This work is based in part on observations made with the Spitzer Space
Telescope, which was operated by the Jet Propulsion Laboratory, California
Institute of Technology under a contract with NASA.  The Spitzer data are
from program ID 11004 (PI: H.  Dole)

%CADE 
We acknowledge the use of data provided by the Centre d'Analyse de Données Etendues (CADE), a service of IRAP-UPS/CNRS (http://cade.irap.omp.eu,~\citep{paradis12}).

%NED
This research has made use of the NASA/IPAC Extragalactic Database, which is funded by the National Aeronautics and Space Administration and operated by the California Institute of Technology.

%Software
{\it Software:} This research made use of astropy, a community developed
core Python package for astronomy \citep{astropy}, APLpy, an open-source
plotting package for Python \citep{aplpy}, the IDL Astronomy Library
\citep{Landsman1993}, \texttt{CIGALE} \citep{boquien19}, the \texttt{SWarp}
program \citep{swarp}, Trilogy \citep{trilogy} and TOPCAT
(http://www.starlink.ac.uk/topcat/).
This research has made use of the VizieR catalog access tool, CDS,
Strasbourg, France \citep{10.26093/cds/vizier}. The original description 
of the VizieR service was published in \citet{vizier2000}.

%Facilities
{\it Facilities:} Mayall, Bok, WISE, NEOWISE, \herschel, \spitzer, GALEX,
NED, VLA, LOFAR, WSRT.

\end{acknowledgements}

\begin{appendix}
\onecolumn
    
\section{Spectral energy distributions of group galaxies}\label{app:seds}

\begin{figure*}[h!]
        \centering
        \includegraphics[width=\hsize]{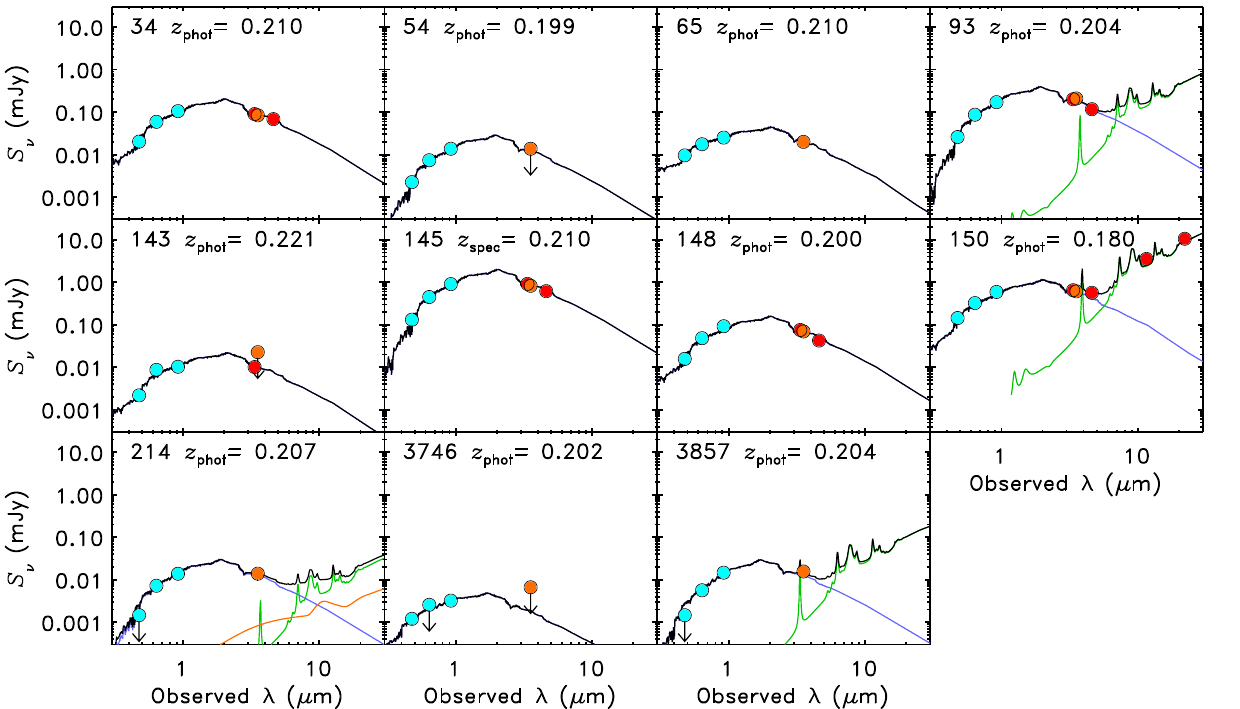}
        \caption{SEDs of the group members in the J1248 field (filled
        circles, DESI in cyan, IRAC in orange, and WISE in red). The downward arrows represent 5$\sigma$ upper limits. The \texttt{CIGALE} best-fit model (black line) and the model components
        (stars: blue; dust: green; AGN: orange) are also shown. The source
        IDs and redshifts are annotated.}
        \label{fig:seds}
    \end{figure*}

The field of J1248 contains a galaxy group at $z=0.2$ initially identified using DESI photometric and spectroscopic redshifts by \citet{wen24}. There are 11 members in the field of which six are newly identified through our photometric redshifts. All members have redshifts between 0.18 and 0.22, equivalent to  |$\Delta v$|$\lesssim 5000$\,\kms. The SEDs of the group galaxies in the field have been modeled with \texttt{CIGALE}. The derived main properties are listed in Table~\ref{tab:group}, and their SEDs and best-fit models are shown in Fig.~\ref{fig:seds}.\null

\begin{table*}[h!]
\caption{\label{tab:group}Group galaxies in the J1248$+$4826 field.}
\centering
\renewcommand{\arraystretch}{1.2}
\begin{tabular}{rc cc ccc}
\hline\hline
   ID   &  DESI\tablefootmark{a}&$\alpha$\tablefootmark{a}&$\delta$\tablefootmark{a}& $z_{\rm phot}$\tablefootmark{b} & M$_{\rm star}$\tablefootmark{b} & SFR\tablefootmark{b} \\
        &   ID   &     (deg)     &      (deg)    &                      &   (10$^{10}$\msun)     &  (\msun\,yr$^{-1}$)   \\
\hline
      34  &  1878  &  192.193934 &  48.423785  &  0.2052$\pm$0.0254\tablefootmark{c}  &   1.96$\pm$0.32  &   0.57$\pm$0.16  \\
      54  &  1831  &  192.190929 &  48.428811  &  0.199$\pm$0.094                     &   0.33$\pm$0.24  &   0.14$\pm$0.27  \\
      65  &  2303  &  192.232948 &  48.432341  &  0.2096$\pm$0.0631\tablefootmark{d}  &   0.30$\pm$0.08  &   0.46$\pm$0.11  \\
      93  &  2056  &  192.209441 &  48.443826  &  0.204$\pm$0.096                     &   4.81$\pm$2.84  &   0.81$\pm$1.55  \\
     143  &  1928  &  192.198684 &  48.432839  &  0.221$\pm$0.093                     &   0.23$\pm$0.14  &   0.21$\pm$0.27  \\
     145  &  2126  &  192.216211 &  48.437844  &  0.2081\tablefootmark{e}             &  25.63$\pm$1.28  &   0.50$\pm$0.55  \\
     148  &  2211  &  192.224537 &  48.438518  &  0.2039$\pm$0.0221\tablefootmark{c}  &   1.57$\pm$0.20  &   0.40$\pm$0.06  \\
     150  &  2280  &  192.230275 &  48.435038  &  0.1831$\pm$0.0190\tablefootmark{c}  &   6.89$\pm$1.14  &  10.86$\pm$1.02  \\
     214  &  2125  &  192.216206 &  48.426922  &  0.207$\pm$0.141                     &   0.35$\pm$0.35  &   0.47$\pm$1.93  \\
    3746  &  1876  &  192.193608 &  48.425098  &  0.202$\pm$0.177                     &   0.59$\pm$0.09  &   0.22$\pm$0.81  \\
    3857  &  2058  &  192.209723 &  48.434673  &  0.204$\pm$0.119                     &   0.44$\pm$0.35  &   0.21$\pm$0.70  \\
\hline
\hline
\end{tabular}\\
\tablefoot{
\tablefoottext{a}{\small Object identifiers (DESI ID) and coordinates
($\alpha$, and $\delta$) were derived from the DESI DR10 catalog.}
\tablefoottext{b}{\small Parameters obtained from modeling the SED with
\texttt{CIGALE} \citep{boquien19}. }
\tablefoottext{c}{\small DESI photometric redshift from \citet{wen24}.}
\tablefoottext{d}{\small DESI photometric redshift from \citet{zou22}.}
\tablefoottext{e}{\small Spectroscopic redshift from SDSS DR6 \citep{sdss_dr6}.}
}
\end{table*}

\section{The heterogeneity of sources designated as ORCs}\label{app:orcs}

To contextualize the classification challenges presented by J1248$+$4826, which we initially considered an ORC, we summarize here the growing diversity of sources designated as ORCs and discuss the resulting ambiguity in the use of this term. The canonical definition of an ORC is a limb-brightened radio ring, typically spanning about one arc minute across, centered on a massive (several times $10^{11}$\,\msun) ETG at $z \sim 0.2 - 0.6$ \citep{norris25}. However, as the search for ORCs has expanded through increasingly sensitive sub-gigahertz radio observations, the literature has come to include a broader variety of sources that depart from this strict definition. Several designated ORCs lack the characteristic edge-brightened structure or a central elliptical host, instead appearing as diffuse circular sources with a complex internal structure \citep[e.g.,][]{kumari24,koribalski24_physalis,bulbul24}.\null

An up-to-date list of sources designated as ORCs in the literature, together with their main properties, is reported in Table~\ref{tab:lit_orcs}. We divide these into three broad categories: (i) multiple nested radio shells surrounding a brightest group galaxy (BGG; i.e., Cloverleaf, \citealt{bulbul24}; and Physalis, \citealt{koribalski24_physalis}), (ii) the classical limb-brightened ORCs \citep{norris21_orc_discovery,koribalski21,norris25,koribalski24,koribalski26}, and (iii) those discovered in LOFAR surveys \citep{omar22,hota25,degasperin26}. We also include a fourth category of diffuse circular radio sources, although these are not always explicitly classified as ORCs \citep{kumari24,kumari24_horseshoe}. \null

\begin{table*}
\caption{\label{tab:lit_orcs}Properties of ORCs and ORC-like sources from the literature.}
\centering
\renewcommand{\arraystretch}{1.2}
\setlength{\tabcolsep}{4pt}
\begin{tabular}{lccccccc}
\hline\hline
Source name                           & Redshift& Radius\tablefootmark{a} & Radio      & Host\tablefootmark{b}       & log(P$_{\rm 150}$) & log(M$_{\rm halo}$)\tablefootmark{c} & Ref.\tablefootmark{d}\\
                                      &         & (kpc)  & Morphology & Properties & (W\,Hz$^{-1}$)   &   (\msun)       &     \\
\hline
J1248$+$4826\tablefootmark{e}         & 0.20    & \030   & HSS+Diffuse & R\tablefootmark{f}, O(16\arcsec), Gr & 24.78$\pm$0.02  & 13.58--13.78   & (1) \\ 
\hline
\multicolumn{8}{c}{Nested shells around a BGG}\\
Physalis/J1914$-$5433                 & 0.017   & \073   & Double Shells& R, C, pair, Gr & 23.70$\pm$0.01  & 12.78--13.30  & (2) \\ 
Cloverleaf/J1137$-$0050               & 0.046   & \090   & Shells+Diffuse & R, C, Gr     & 25.11$\pm$0.02  & 12.78--13.30  & (3) \\ 
\hline
\multicolumn{8}{c}{Limb-brightened ORCs}\\
\object{ORC1}/J2103–6200              & 0.551   & 260    & Ring         &  C           & 25.71$\pm$0.07  & 14.42--14.58  &   (4) \\ 
ORC2\&3/J2058$-$5736\tablefootmark{g} & 0.252 & 163    & Double Ring+Diffuse & R, O (15\arcsec) & 24.64$\pm$0.20  & \nodata       &   (4) \\ %R, O (37\arcsec)
\object{ORC4}/J1555+2726              & 0.4515  & 180    & Double Ring  & R, C         & 25.48$\pm$0.12  & 13.20--13.40  &   (4,5,6) \\ %or double
\object{ORC5}/J0102$-$2450            & 0.270   & 150    & Ring         & R, C         & 24.64$\pm$0.03  & 12.38--12.53  &   (7) \\ 
ORC6\tablefootmark{h}/J1841$-$6547    & 0.18  & 365 & Double ring & R, C         & 26.81$\pm$0.01  & \nodata       &   (8) \\ 
MIGHTEE/\object{ORC J0219$-$0505}     & 0.196   &\057    & Ring         &  R, C        & 24.20$\pm$0.05  & \nodata       &  (9) \\ 
\object{ORC J0356$-$4216}             & 0.494   & 188    & Ring         &  R, C        & 25.85$\pm$0.03  & \nodata       &  (10) \\ 
\object{ORC J1027$-$4422}             &$\sim$0.3&  200   & Ring         & R, C, Gr     & $<$24.70         & 11.71--11.72 &  (11) \\
ORC J0210$-$5710                      & 0.382   & 219    & Ring         &  R, C        & 25.27$\pm$0.06  & \nodata       &  (12) \\ 
ORC J0402$-$5321                      & 0.548   & 130    & Ring         &  R, O(12\arcsec) & 25.29$\pm$0.03  & \nodata   &  (12) \\ 
ORC J0452$-$6231                      & 0.419   & 150    & Ring         &  C           & 25.13$\pm$0.03  & \nodata       &  (12) \\ 
ORC J1313$-$4709                      & \nodata & 30\arcsec & Ring      &  R, C        & \nodata         & \nodata       &  (12) \\ 
ORC J2304$-$7129                      & 0.148   &\080    & Ring         &  R, C        & 24.30$\pm$0.03  & \nodata       &  (12) \\ 
\hline
\multicolumn{8}{c}{LOFAR ORCs}\\
\object{J0020+3018}                   & \nodata & 27\arcsec & Ring      &  unidentified        & \nodata         & \nodata      &  (13) \\ 
\object{ORC J0823+6216}               & 0.039   &\044    & Clumpy ring  & R, O(25\arcsec), Gr & 23.32$\pm$0.01  & \nodata   &  (14) \\ 
\object{ORC J0825+6132}               & 0.800   & 136    & Clumpy double ring  & C     & 25.44$\pm$0.03  & \nodata      &  (14) \\ 
\object{ORC J0847+7022}               & 0.405   & 246    & Double ring  & R, C, Gr     & 25.54$\pm$0.01  & \nodata      &  (14) \\ % partial and filled
\object{ORC J1134+6642}               & 0.400   & 339    & Clumpy ring  & R, O(13\arcsec) & 25.37$\pm$0.02  & \nodata   &  (14) \\ 
\object{ORC J1150+2449}               & 0.670   & 127    & Clumpy ring  & O(3\arcsec) & 25.18$\pm$0.04  & \nodata      &  (14) \\ 
\object{ORC J1222+6436}               & 0.245   & 112    & Partial ring & C            & 23.99$\pm$0.04  & \nodata      &  (14) \\ 
\object{ORC J1313+5003}               & 0.860   & 142    & Double  ring & R, C         & 26.16$\pm$0.05  & \nodata      &  (14,15) \\ 
Uruk-hai/\object{ORC J1633+1441}      & 0.128   & 208    & Partial ring & R, C, Gr     & 24.52$\pm$0.02  & \nodata      &  (14) \\ 
\hline
\multicolumn{8}{c}{Circular diffuse radio sources}\\
J1218+1813                            & 0.14    &\090    & Diffuse      & R, O (4\arcsec)  & 25.76$\pm$0.02  & \nodata       &  (16) \\ %diffuse
J1407+0453\tablefootmark{i}           &0.134 &\080 &Diffuse+HSS & R, O (5.5\arcsec)& 25.57           & \nodata       &  (17) \\
J1507+3013                            & 0.079   &\068    & Diffuse      & R, O (6\arcsec)  & 25.56$\pm$0.01  & \nodata       &  (18) \\ %diffuse
\hline
\hline
\end{tabular}\\
\tablefoot{
\tablefoottext{a}{Ring radius. For the double-headed ORCs, the size refers to the average of the two ring radii. } 
\tablefoottext{b}{R: Radio detected host; C: Central; O: Offset from ORC center with approximate value given in parenthesis; Gr: Associated with a galaxy group. For the double-headed ORCs, the host is considered central if located in between the two rings.}
\tablefoottext{c}{Halo masses refer to M$_{\rm 500}$ and were derived from \citet{koribalski24_physalis}.} 
\tablefoottext{d}{References: (1) This work;  (2) \citet{koribalski24_physalis}; (3) \citet{bulbul24}; (4) \citet{norris21_orc_discovery}; (5) \citet{coil24}; (6) \citet{coil26}; (7) \citet{koribalski21}; (8) \citet{koribalski26}; (9) \citet{norris25}; (10) \citet{taziaux25}; (11) \citet{koribalski24}; (12) \citet{gupta25}; (13) \cite{omar22}; (14) \citet{degasperin26}; (15) \citet{hota25}; (16) \citet{kumari25}; (17) \citet{kumari24_horseshoe}; (18) \citet{kumari24}.} 
\tablefoottext{e}{Radio emission associated with LoTSS IDs 85256, 16906, 16908, and 16909 (see Table~\ref{tab:lofar_data}).}
\tablefoottext{f}{We refer to ID 145 as the host of J1248. This galaxy is not individually detected in LOFAR. Rather, it is part of LoTSS ID 16908 and associated with a bright blob in the LoLSS image at 54\,MHz. Therefore, it is considered a radio faint source with an ultra-steep radio spectral index.}
\tablefoottext{g}{The ORC2\&3 system is no longer considered an ORC but rather the lobes of a restarted radio galaxy whose identification remains uncertain. Here, we adopt the galaxy labeled "C"-- as in \citet{norris21_orc_discovery} following \citet{rupke24} -- as the host.}
\tablefoottext{h}{We detect two different ORCs in the literature named ORC6. In this work, ORC6 refers to ORC\,J1841-6547, as in \citet{koribalski26}. Howeover, in \citet{kumari24} the same name is assigned to ORC\,J0020$+$3018 \citep{omar22}.} 
\tablefoottext{i}{J1407$+$0453 is not designated as ORC.}
}
\end{table*}

The diversity of sources included in the compilation suggests either that the term ORC encompasses a heterogeneous mixture of distinct astrophysical phenomena or that the boundaries of the ORC class remain insufficiently well defined. \citet{norris25} proposed the following strict classification: edge-brightened circles of radio emission, with or without internal structures, that are centered on a distant central galaxy and that do not have multiwavelength counterparts. The observed ring results from a violent event, either starburst- or AGN-related, that re-energizes a spherical shell, bubble, or remnant radio lobe of previously ($\sim$1\,Gyr ago) deposited magnetized plasma  \citep{norris22,fujita24}. This interpretation is supported by the detection of shocked ionized gas in ORC4 \citep{coil24,coil26}, and by spectroscopic observations of the central galaxy of other ORCs, which are consistent with old galaxies with bursty pasts at their center \citep{rupke24}.\null

Applying this strict definition would, however, require the reclassification of sources that lack well defined rings, have offset hosts, or contain extended diffuse components. Some of these systems may result from different formation mechanisms, in which the energy injection is driven primarily by large-scale disturbances in the surrounding medium. Examples include merging galaxy groups \citep{shabala24,wang26}, galaxy mergers \citep{dolag23}, and accreting gas \citep{yamasaki24}, all of which can produce ring-like radio sources. These large-scale processes may nevertheless  trigger an energetic event in a member galaxy, thereby intertwining the effects of external disturbances and host activity. This could explain the difficulty of disentangling ORCs generated primarily by the activity of an individual galaxy from those resulting from large-scale dynamics.\null

It is therefore plausible that some of the sources listed in Table~\ref{tab:lit_orcs}, represent a class of diffuse radio structures distinct from the classical ORCs. Alternatively, these sources may share a common evolutionary connection with classical ORCs, circular diffuse radio sources, and remnant radio lobes, but be observed at different evolutionary stages or under different dynamical conditions. Distinguishing between these possibilities will require larger samples and spatially resolved radio, optical, and X-ray observations.
\twocolumn
\end{appendix}
\end{document}